\documentclass[fleqn,usenatbib,usedcolumn]{mnras}

\usepackage[T1]{fontenc}
\usepackage{ae,aecompl}

\usepackage{graphicx}	
\usepackage{amsmath}	
\usepackage{amssymb}	

\DeclareRobustCommand{\VAN}[3]{#2}
\let\VANthebibliography\thebibliography
\def\thebibliography{\DeclareRobustCommand{\VAN}[3]{##3}\VANthebibliography}

\title[Planetary migration in precessing disks]{Planetary migration in precessing disks for S-type wide binaries}

\author[A. Roisin et al.]{
Arnaud Roisin,$^{1}$\thanks{E-mail: arnaud.roisin@unamur.be}
Jean Teyssandier,$^{1}$
and Anne-Sophie Libert$^{1}$
\\
$^{1}$naXys, Department of Mathematics, University of Namur, 61 Rue de Bruxelles, B-5000 Namur, Belgium
}

\date{Accepted XXX. Received YYY; in original form ZZZ}

\pubyear{2021}

\begin{document}
\label{firstpage}
\pagerange{\pageref{firstpage}--\pageref{lastpage}}
\maketitle

\begin{abstract}
The discovery of numerous circumprimary planets in the last few years has brought to the fore the question of planet formation in binary systems. The significant dynamical influence, during the protoplanetary disk phase, of a binary companion on a giant planet has previously been highlighted for wide binary stars. In particular, highly inclined binary companion can induce perturbations on the disk and the planets, through the Lidov-Kozai resonance, which could inhibit the formation process. In this work, we aim to study how the disk gravitational potential acting on the planet and the nodal precession induced by the wide binary companion with separation of 1000 AU on the disk act to suppress the Lidov-Kozai perturbations on a migrating giant planet. We derive new approximate formulas for the evolution of the disk's inclination and longitude of the ascending node, in the case of a rigidly precessing disk with a decreasing mass and perturbed by a wide binary companion, which are suitable for $N$-body simulations. We carry out 3200 simulations with several eccentricity and inclination values for the binary companion. The gravitational and damping forces exerted by the disk on the planet tend to keep the latter in the midplane of the former, and suppress the effect of the binary companion by preventing the planet from getting locked in the Lidov-Kozai resonance during the disk phase. We also confirm that because of nodal precession induced by the binary, a primordial spin-orbit misalignment could be generated for circumprimary planets with an inclined binary companion. 

\end{abstract}




\section{Introduction}
More than 100 S-type planets (or circumprimary planets) have been discovered \citep{Schwarz_2016} in multiple-star systems, most of them found in wide binaries with separations of at least 500 AU. Recent studies highlighted the different parametric distributions of planets around binary stars compared to single stars, in particular differences in semi-major axis distributions \citep{Hirsch_2020} and slightly higher eccentricities for giant planets in wide binaries \citep{Kaib_2013}. These observations lead to the question of the influence that could have a binary companion on the formation of planets in binary systems as well as on their long-term evolution and stability.

It has previously been shown that a binary companion could have a strong impact on the dynamics of S-type planets, even leading to retrograde orbits \citep{Naoz_2013_b} as well as strong variations in eccentricities through the Lidov-Kozai resonance \citep{Kozai_1962, Lidov_1962}. Different studies have been carried out to investigate the stability and habitability of planets in binary star systems for some specific observed giant planet systems \citep[e.g.,][]{Dvorak_2003,Haghighipour_2006,Bazso_2017, Giuppone_2017} or more generally to determine stable and habitable zones for different parametric choices \citep[e.g.,][]{Holman_1999,Eggl_2014,Marzari_2016,Quarles_2020}. Most of the studies focused on close binary systems due to the strong gravitational interactions exerted by the binary companion. Nevertheless, the influence of a wide binary companion could also be significant during the formation of a planetary system.

During the disk phase, the binary companion exerts influence over the disk around the primary. Among others, it affects the formation of the disk \citep{Franchini_2019,Chachan_2019}, its truncation \citep{Artymowicz_1994} and its shape \citep{Terquem_1993,Papaloizou_1995}. Lidov-Kozai cycles caused by the binary companion can also be observed within the disk \citep{Martin_2014, Fu_2015a}, but they tend to disappear when considering the self-gravity in massive circular disks \citep{Batygin_2011,Rafikov_2015_b}. The companion also influences the stability of the disk and Lidov-Kozai instabilities are possible for sufficiently inclined disks \citep{Lubow_2017,Zanazzi_2017}. Moreover, hydrodynamical simulations of the evolution of a giant planet in the disk have been carried out \citep[e.g.,][]{Xiang_2014, Picogna_2015,Lubow_2016,Martin_2016}, where they observed that a misalignment generally occurs between the disk and the orbit of the planet in the presence of a highly inclined binary companion.

In a previous study \citep{Roisin_2020}, we discussed the influence of a wide binary companion on the evolution of a giant planet migrating in the disk. This work was conducted using a symplectic $N$-body integrator adapted for binary star systems, in which the dissipation due to the migration in the disk was modelled by appropriate formulae fitted from hydrodynamical simulations by \citet{Bitsch_2013}. We showed that a capture of the planet in a Lidov-Kozai resonant state is not automatic when the binary companion is highly inclined, as a locking in the resonance happens in approximately one third of the cases only. This preliminary work neglected two potentially significant effects related to the disk, that is the disk gravitational potential acting on the planet and the nodal precession induced by the binary companion on the disk. The purpose of the present work is to study the influence of these effects on the previous results, in particular on the capture of the systems in the Lidov-Kozai resonance. 

The gravitational potential exerted by a disk on an inclined planet was previously studied for a flat disk in \cite{Terquem_2010}. They reported Lidov-Kozai cycles (associated with high values of the eccentricity) for a planet whose inclination is smaller than the critical angle value identified for an outer perturber in \cite{Lidov_1962,Kozai_1962}. The effects of disk warping on the planetary inclination was considered in \cite{Terquem_2013}, where it was shown that the gravitational potential in 3D wrapped disks generally leads to the precession of the planet and could in some cases increase the planetary eccentricity. \cite{Teyssandier_2013} considered frictional forces from 3D disk leading to eccentricity and inclination damping in addition to disk gravitational potential for inclined planets. They observed Lidov-Kozai cycles exerted by the disk on the planet for inclination value as small as 20$^\circ$.

The precession of the disk caused by a binary companion was studied in \cite{Batygin_2011}, using both a Gauss's averaging method and an alternative method based on the Laplace-Lagrange secular theory, following the idea from \cite{Levison_2007}, and consisting in dividing the disk in several annuli. Further studies where carried out combining various effects, such as accretion, magnetic star-disk interaction and solar winds, to explain the misalignment between the stellar spin axis and the disk angular momentum \citep[see e.g.][]{Batygin_2013,Lai_2014,Spalding_2014}. The nodal precession was also studied including disk warping in \citet{Zanazzi_2018a} and a migrating planet in \citet{Zanazzi_2018b}.

When the disk mass decreases exponentially with time, as it is the case here, the analytical solution of \citet{Murray_1999} and \citet{Batygin_2011} for the disk nodal precession is not convenient and we have developed here new approximate formulas for the evolution of the disk's inclination and longitude of the ascending node suitable for a rigidly precessing disk perturbed by a wide binary companion.

The paper is organized as follows. In Sect.~\ref{Simulation}, we describe the $N$-body code used for the simulations and derive formulas for the disk gravitational potential acting on the planet and the nodal precession induced by the binary companion on the disk. Sect.~\ref{results} presents the results of the simulations by emphasing the influence of the two newly added effects on the evolution of the planet during the disk phase. Finally, our results are summarized in Sect.~\ref{conclusion}.

\section{Code overview and numerical methods}
\label{Simulation}
In \citet{Roisin_2020}, we studied the influence of a wide binary companion on the evolution of a migrating giant S-type planet embedded in a protoplanetary disk. In the present work, we focus on two additional effects that were neglected in \citet{Roisin_2020}, namely the disk gravitational potential acting on the giant planet and the nodal precession induced by the binary companion on the disk.

\subsection{$N$-body code for migrating planets in binaries}\label{sec:nbody}
The study is carried out with the symplectic $N$-body code adapted for migrating planets in binary star systems and presented in \citet{Roisin_2020} to which we refer for more details. 
The code consists in a modification of the well-known SyMBA integrator \citep{Duncan_1998} in order to include an outer binary companion by following \citet{Chambers_2002}. The Type-II migration of giant planets in a protoplanetary disk is also included, as well as damping forces. In the formulas that we used for migration and damping, derived through three-dimensional hydrodynamical simulations in \citet{Bitsch_2013}, a gap forms in the surface density profile in the disc, even for inclined planets. In particular, this means that the damping forces are reduced by the presence of the gap. Because the damping formulas derived from these simulations use the perturbed surface density and not the unperturbed one, no artificial damping is added. The timescale adopted for the migration corresponds to the viscous accretion timescale multiplied by the ratio of the planetary mass over the local disk mass when the mass of the planet is comparable to the mass of the material in its vicinity \citep{Ivanov_1999,Nelson_2000,Crida_2007}. In order to include the dissipation on the planetary eccentricity and inclination due to the disk as a force in the N -body code, the timescales from \citet{Bitsch_2013} were used through the acceleration formulas from \citet{Papaloizou_2000} and implemented in the code following \cite{Lee_2002} to preserve the symmetry of the symplectic algorithm. Moreover, the mass of the disk is decreased exponentially through the evolution of the system. The disk can be considered as dissipated and thus neglected in the code when
\begin{equation}
\frac{d{m}_{\rm disk}(t)}{dt}=\frac{m_{\rm disk,0}}{T_0} \exp{(-t/T_0})<10^{-9} m_0/yr,
\end{equation}
where $m_0$ denotes the mass of the central star, $m_{\rm disk,0}$ the disk mass at $t=0$ fixed here to 8 $m_{\rm Jup}$, and $T_0=2.8\times 10^5$ yr which corresponds to a disk dispersal time of $\sim 1$ Myr \citep{Mamajek_2009}. 

We now present the additional effects, newly implemented in the code, whose influence is studied here: the gravitational planet-disk interactions (Section \ref{GP}) and the nodal precession of the binary on the disk (Section \ref{NP}).

\subsection{Disk gravitational potential}\label{GP}

We first start by considering the gravitational potential that a disk exterts on a planet. 
We assume that the disk is flat and  axisymmetric. The potential it exerts on a planet can be written in spherical coordinates as \citep[see, e.g.,][]{Terquem_2010}:
\begin{equation}
	\Phi(r,\varphi,\theta) = -G \int_{a_{in}}^{a_{out}} \Sigma(\tilde{r}) \tilde{r} \int_{0}^{2\pi} \frac{ d\tilde{\varphi} d\tilde{r}}{\sqrt{r^2+\tilde{r}^2-2r\tilde{r}\cos\left(\tilde{\varphi}\right)\sin(\theta)}},
\end{equation}
where $(r,\varphi,\theta)$ correspond to the spherical coordinates of the planet, $\Sigma$ is the disk surface density, $a_{in}$ and $a_{out}$ are the inner and outer edges of the disk and G is the gravitational constant. Following \citet{Hure_2011}, the potential can be expressed as 
\begin{equation}
\Phi(r,\varphi,\theta) = -2G \int_{a_{in}}^{a_{out}} \Sigma(\tilde{r}) \sqrt{\frac{\tilde{r}}{r \sin(\theta)}} k K(k) d\tilde{r},
\end{equation}
where $K(k)$ is the complete elliptic integral of first kind:
\begin{equation}
	K(k)=\int_{0}^{\pi/2} \frac{d\tilde{\varphi}}{\sqrt{1-k^2\sin^2\left(\tilde{\varphi}\right)}}
\end{equation}
with $k$ between 0 and 1 and equal, in our case, to 
\begin{equation}
	k=2\sqrt{\frac{r \tilde{r} \sin(\theta)}{r^2+\tilde{r}^2+2r\tilde{r}\sin(\theta)}}.
\end{equation} 
To implement the potential in the code and preserve the symmetry of the integration scheme, we converted it to an acceleration in Cartesian heliocentric coordinates, which leads to the following expression
\begin{equation}
\begin{aligned}
	\frac{d^2x}{dt^2} &= - \sin(\theta) \cos(\varphi) \frac{\partial \Phi}{\partial r} - \cos(\theta) \cos(\varphi) \frac{1}{r} \frac{\partial \Phi }{\partial \theta} \\
	\frac{d^2y}{dt^2} &= - \sin(\theta) \sin(\varphi) \frac{\partial \Phi}{\partial r} - \cos(\theta) \sin(\varphi) \frac{1}{r} \frac{\partial \Phi }{\partial \theta} \\
	\frac{d^2z}{dt^2} &= - \cos(\theta) \frac{\partial \Phi}{\partial r} + \sin(\theta)  \frac{1}{r} \frac{\partial \Phi }{\partial \theta}, 
\end{aligned}
\end{equation}
where 
\begin{equation}
	\begin{aligned}
		\frac{\partial \Phi}{\partial r} &= \frac{G}{r} \int_{a_{in}}^{a_{out}} \Sigma(\tilde{r}) \sqrt{\frac{\tilde{r}}{r \sin(\theta)}} k \left[K(k) - \frac{\tilde{r}^2-r^2}{R^2}E(k)\right] d\tilde{r} \\
		\frac{\partial \Phi}{\partial \theta} &= G \int_{a_{in}}^{a_{out}} \Sigma(\tilde{r}) \cot(\theta)\sqrt{\frac{\tilde{r}}{r \sin(\theta)}} k \left[K(k) - \frac{\tilde{r}^2+r^2}{R^2}E(k)\right] d\tilde{r}
	\end{aligned}
	\label{derivative}
\end{equation}
with $R^2=r^2+\tilde{r}^2-2r\tilde{r}\sin(\theta)$ and $E(k)$ the complete elliptic integral of the second kind: 
\begin{equation}
	E(k)=\int_{0}^{\pi/2} \sqrt{1-k^2\sin^2\left(\tilde{\varphi}\right)} d\tilde{\varphi}.
\end{equation}
In order to evaluate the complete elliptic integrals, we used the approximation given by \cite{Abramowitz_1972}. The other integrals are calculated using the Romberg's method presented in \cite{Press_1992}. A divergence can arise in Eqs.~\eqref{derivative} when $R=0$, that is, when the planet is exactly in the plane of the disk. In reality, massive planets tend to open a gap in disks, and the gravitational interaction at this location is reduced \citep[even when the planet is on an inclined orbit with respect to the disk's midplane, see e.g.,][]{Bitsch_2013}. In order to prevent $R^2$ from becoming 0, we simply replaced it by $R^2+10^{-6}$. Note that we also tried splitting the disk into two parts, one extending from $a_{in}$ to $r-\Delta_{gap}/2$ and one from $r+\Delta_{gap}/2$ to $a_{out}$ (where $\Delta_{gap}$ represents the gap width). We also tested the effect of removing the component of the potential within the Hill sphere of the planet. Both prescriptions did not significantly alter our results and we kept the simplest prescription mentioned above.

In this section we have evaluated the gravitational force exerted by the disk on the planet. On the other hand, the planet is also exerting a force on the disk. In particular, we are interested in this paper in planets whose orbits can potentially be strongly inclined with respect to the disk's midplane. Such planets could in principle create a warp in the disk. A linear calculation by \citet{Teyssandier_2013} showed that the back-reaction of the planet on the disk exterts only a small torque and can be neglected.

\subsection{Disk nodal precession caused by a binary star companion} \label{NP}

The second effect studied in this work is the disk response to the perturbation from a binary star companion. We describe how the numerical approximation is constructed in Section~\ref{sec:approx}, while an analytical validation is performed in Section~\ref{sec:analytic}. Note that the approximation found here is only appropriate for rigidly precessing disks perturbed by a wide binary companion.

\subsubsection{Construction of a numerical approximation}
\label{sec:approx}

\begin{figure}
	\includegraphics[width=\columnwidth]{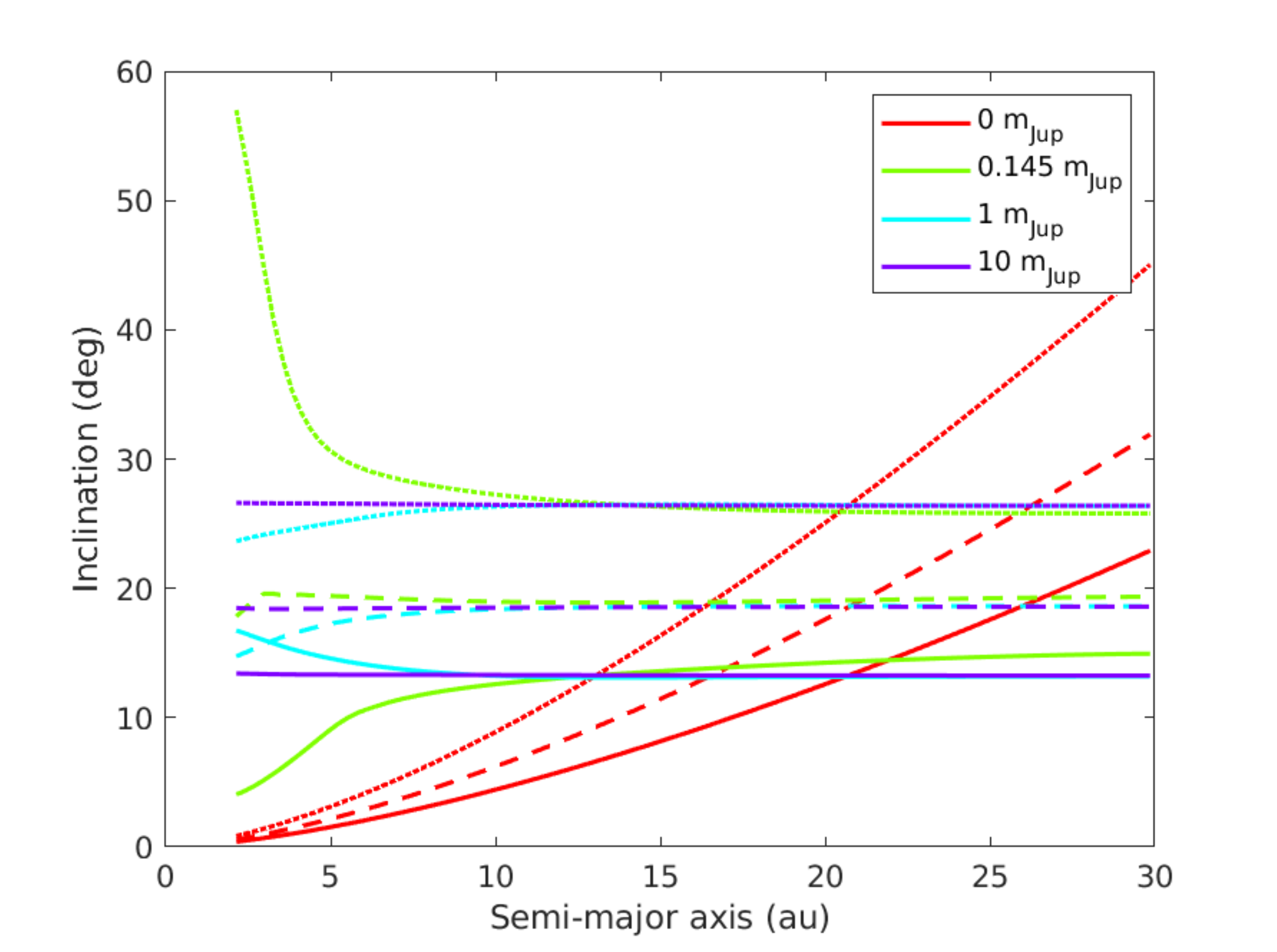}
	\caption{Time evolution of the inclination of the disk as a function of the semi-major axis, for different disk masses (in the initial disk plane reference frame). The color code refers to the different masses and the line type to different times. The solid line stands for t=0.5 Myr, the dashed line for t=0.75 Myr and the dotted line for t=1 Myr. The inclination of the binary $i_B$ is fixed to $60^\circ$.}
	\label{disk_evol_mass}
\end{figure}

Our purpose is to determine formulas for the evolution of the inclination and longitude of the ascending node of the disk.
The formulas deduced here are consistent only if the disk maintains an uniform inclination. This condition can be fulfilled when the disk self-gravity is considered in the case of a wide binary companion with separation of 1000 AU \citep[see][and our Fig.~\ref{disk_evol_mass}]{Batygin_2011}.
The formulas are expressed with respect to the initial plane of the disk.

We followed the method of \cite{Batygin_2011} based on \cite{Levison_2007}. This method consists in splitting the disk in several massive self-gravitating rings adjacent to one another, and modelling their interaction using the classical Laplace-Lagrange secular theory. This assumes that the disk is dominated by its self-gravity, as opposed to internal pressure forces \citep[in that case, see, e.g,][]{Larwood_1996,Zanazzi_2018a}. Note that we suppose that the disk (and thus each individual ring) remains circular and therefore rings do not overlap with each others, preventing orbital crossing \citep[see][]{Batygin_2011}.

The Hamiltonian of a given ring $j$ is expressed by \citep[see][]{Murray_1999}:
\begin{equation}
\mathcal{H}_j = \frac{1}{2} B_{jj} i_j^2 + \sum\limits_{k=1, k\neq j}^{N+1} B_{jk} i_j i_k \cos(\Omega_j-\Omega_k)
\end{equation}
with $k$ being the indice of the different rings (note that we consider the binary companion as the $N+1$-th ring), $i_k$ the inclination and $\Omega_k$ the longitude of the node of the $k$-th ring. The coefficients are given by
\begin{align}
	B_{jj} &= -\frac{n_j}{4} \sum\limits_{k=1, k\neq j}^{N+1} \frac{m_k}{m_0+m_j} \alpha_{jk} \bar{\alpha}_{jk} b^{(1)}_{3/2}(\alpha_{jk}) \\
	B_{jk} &= \frac{n_j}{4} \frac{m_k}{m_0+m_j} \alpha_{jk} \bar{\alpha}_{jk} b^{(1)}_{3/2}(\alpha_{jk})
\end{align}
with $n_j$ the mean motion of $j$-th ring, $m_j$ the mass of the $j$-th ring, $m_0$ the mass of the central star, and $b^{(1)}_{3/2}$ the Laplace coefficient of the first kind. Note that $\bar{\alpha}_{jk}$ is equal to $\alpha_{jk}$ if the perturbation is external and to 1 otherwise.

\begin{figure}
	\includegraphics[width=\columnwidth]{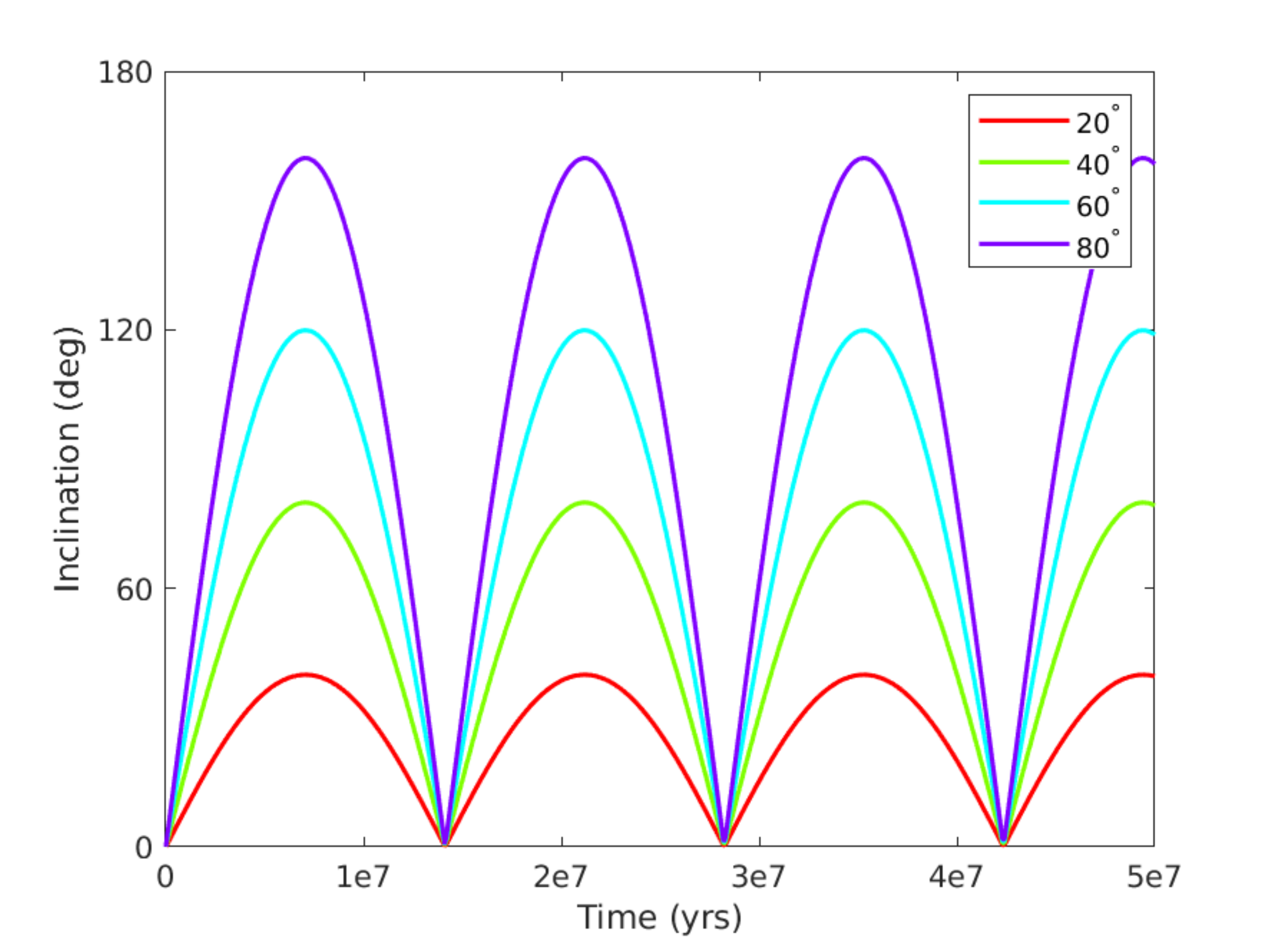}
	\caption{Inclination of the mid-point ring of the disk as a function of time for various inclinations of the binary companion $i_B$. All the inclinations are expressed with respect to the initial plane of the disk.}
	\label{i_evolution_i_B}
\end{figure}

\begin{figure}
	\includegraphics[width=\columnwidth]{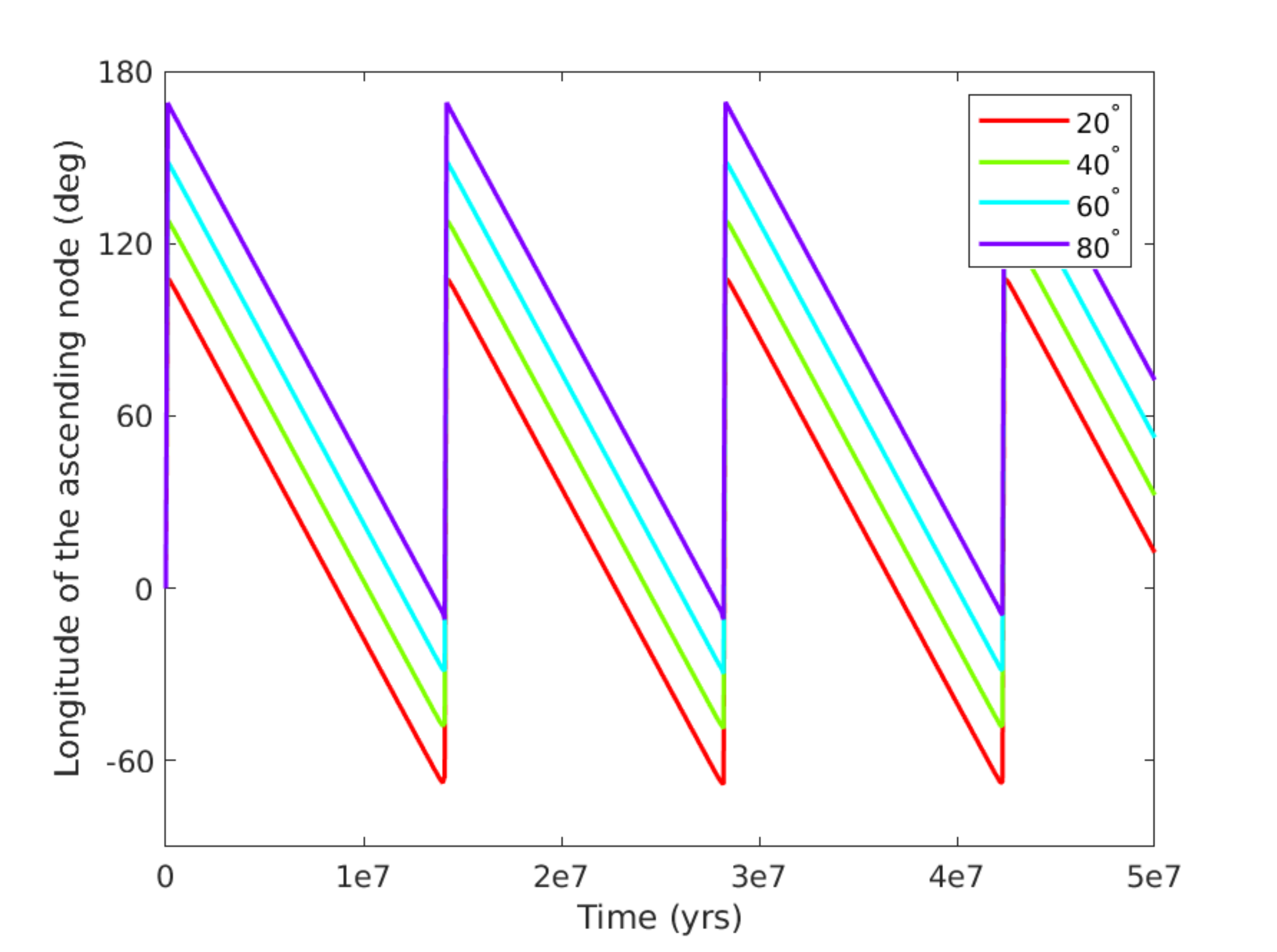}
	\caption{Longitude of the ascending node of the mid-point ring of the disk as a function of time for various longitudes of the ascending node of the binary companion.}
	\label{Om_evolution_i_B}
\end{figure}

To simplify the resolution, \cite{Murray_1999} proposed the following change of coordinates 
\begin{align}
	\label{eq:pq}
	q_j &= i_j \cos(\Omega_j) \nonumber\\
	p_j &= i_j \sin(\Omega_j)
\end{align}
and the Hamiltonian can then be rewritten 
\begin{equation}
\mathcal{H}_j = \frac{1}{2}B_{jj}\left(p_j^2+q_j^2\right) + \sum\limits_{k=1, k\neq j}^{N+1} B_{jk} (p_j p_k + q_j q_k).
\label{Hamiltonian}
\end{equation}

The evolution of the different rings is obtained by solving the Hamiltonian equations. For a disk with a constant mass, an analytical solution can be found in \cite{Murray_1999} and \cite{Batygin_2011}. However, for a decreasing mass model of the disk, as it is the case here, the use of the analytical solution is inappropriate, since the matrix $B$ evolves with time and should be evaluated at each step, which is very time-consuming. For this reason, we aimed to derive approximate formulas for the evolution of the disk's inclination and longitude of the ascending node when perturbed by a binary star.

To do so, we ran several simulations at constant disk mass. Our fiducial case used as a point of comparison is a system with a binary star consisting of two Sun-like stars with a separation of 1000~au and an inclination with respect to the initial plane of the disk $i_B(=i_{N+1})$ of $60^\circ$, a disk with a constant mass of 10 $m_{\rm Jup}$, an inner edge at 2~au, an outer edge at 30~au and a disk surface density proportional to $r^{-0.5}$. Note that this choice of edges is arbitrary. However, \cite{Bate_2018} showed that discs with radial extents of 30 au are found in both simulations and observations. Although such a size is on the smaller side of the distribution in \cite{Bate_2018}, we chose it because it allows our disc to precess coherently under  self-gravity.

Firstly, we tested the influence of the disk mass. Ten different masses for the disk were considered, namely $0,0.145,0.3,0.5,0.7,1,5,8,10,20$ $m_{\rm Jup}$\footnote{In our simulations, $0.145$ $m_{\rm Jup}$ is the disk mass at which the disk is considered to be fully dissipated.}. Fig.~\ref{disk_evol_mass} displays the state of the disk at three different times (0.5, 0.75, and 1 Myr) for four different disk masses. When the disk is massive, there is no visible influence of the disk mass on the evolution of the inclination and the longitude of the ascending node and we conclude that for our chosen range of parameters, self-gravity efficiently maintains rigid precession throughout the disk. As seen in \cite{Batygin_2011} and on Fig.~\ref{disk_evol_mass}, when the disk has no mass, its rigidity is not preserved since there is no self-gravity to allow communication between the different rings of the disk. Note that when the disk mass is small, the disk has some non-uniform motion close to the inner edge. Hydrodynamical effects (radial pressure force and viscous diffusion) that were not considered here can help the disk to maintain the uniform motion as discussed for instance in \citet{Papaloizou_1995} and \citet{Larwood_1996}. From now, we will therefore consider the disk as having an uniform motion and display all results for the ring that is at a radius half-way between the inner and outer edges of the disk.

\begin{table}
	\centering
	\caption{Parameters set for the determination of $f_d$.}
	\begin{tabular}{l|r}
		\hline
		Parameter & value \\
		\hline
		$a_B$ (au) & 500, 1000, 2000  \\
		$a_{\rm in}$ (au) & 0.1, 0.5, 2  \\
		$a_{\rm out}$ (au) &  20, 30, 40 \\
		$m_B$ ($m_\odot$) & 0.5, 1, 2 \\
		\hline
	\end{tabular}

	\label{param_f}
\end{table}

Secondly, we tested the influence of the inclination of the binary companion $i_B$ by considering four different values: $20, 40, 60$ and $80^\circ$. We used a constant disk mass model and carried on the evolution up to $5\times10^7$ yr. We show on Fig.~\ref{i_evolution_i_B} the time evolution of the inclination of the mid-point ring of the disk. Since the disk precesses around the total angular momentum of the system (which is very close to the binary angular momentum vector), the disk inclination varies between 0 and twice the initial inclination of the binary with respect to the initial disk plane. This explains why the inclination in Fig.~\ref{i_evolution_i_B} follows the absolute value of a sinus with an amplitude equal to twice the inclination of the binary companion (see Appendix~\ref{Calculation_details} for details). Our interpolated formula for the inclination of the disk with respect to the initial plane of the disk $i_d$ will therefore have the form 
\begin{equation}
	i_d(t)=2 i_B\left|\sin(2\pi f_d t)\right|,
	\label{i_d}
\end{equation}
with $f_d$ the frequency of the sinusoidal function.

Thirdly, we run additional simulations with the longitude of the ascending node of the binary companion $\Omega_B$ set to $20, 40, 60$, and $80^\circ$. We observe in Fig.~\ref{Om_evolution_i_B} that the maximum value of $\Omega_d$ is $\Omega_B+\frac{\pi}{2}$ in radians. Note that this value can also be deduced from the expression of the mutual inclination $i_{\rm mut}$ between the disk and the binary companion, rewritten as
$	\Omega_d(t)=\arccos\left(\frac{\cos(i_{\rm mut})-\cos(i_B)\cos(i_d(t))}{\sin(i_d(t))\sin(i_B)}\right)+\Omega_B$
and evaluated at $t=0$. Since, in t=0, we have $i_d=0$, $i_B=i_{mut}$ and $\lim\limits_{t\rightarrow0} \frac{1-\cos\left(i_d(t)\right)}{\sin\left(i_d(t)\right)}=0 $, it can easily be shown that $\lim\limits_{t\rightarrow0} \Omega_d(t) = \frac{\pi}{2} + \Omega_B$.

We also notice in Fig.~\ref{Om_evolution_i_B} the linearly decreasing evolution of the longitude of the ascending node of the disk with a period equal to the one of the inclination of the disk. Our formula for the longitude of the ascending node of the disk in the initial disk plane reference frame can therefore be written in the following way:
\begin{equation}
	\Omega_d(t)=\frac{\pi}{2} + \Omega_B - \left( 2 \pi f_d t \mod \pi\right).
	\label{Om_d}
\end{equation}
The modulo represents the variation of $180^\circ$ for the longitude of the ascending node observed in Fig.~\ref{Om_evolution_i_B}. This restriction comes from the choice of the reference frame and is further detailed in Appendix~\ref{Calculation_details}. A more natural choice of reference frame would be the plane of the binary companion, which almost coincides with the invariant Laplace plane. However, the damping formulas for the eccentricities and inclinations of the planets migrating in the disk from \cite{Bitsch_2013} have been designed for the initial disk plane reference frame, which explains this choice.

We now need to determine the expression of the precession frequency $f_d$. A parametric study was carried out to investigate the influence on the frequency $f_d$ of different parameters: the mass of the binary $m_B$ and the semi-major axes of the binary $a_B$, the inner edge $a_{\rm in}$, and the outer edge $a_{\rm out}$.   The values considered in the study are displayed in Table~\ref{param_f}. Note that we did not notice any impact of the inner edge of the disk on the frequency $f_d$. From this parametric study, we found that the best approximation for $f_d$ is given by 
\begin{equation}
f_d = 0.2145 \left(\frac{a_{\rm out}}{1 \rm au}\right)^{3/2} \left(\frac{a_B}{1 \rm au}\right)^{-3} \left(\frac{m_B}{1 m_\odot}\right).
\label{freq_disk}
\end{equation}
As we will see in Section \ref{sec:analytic}, this scaling with the different relevant quantities is a consequence of the secular interactions and can also be derived analytically.



\begin{figure}
	\includegraphics[width=\columnwidth]{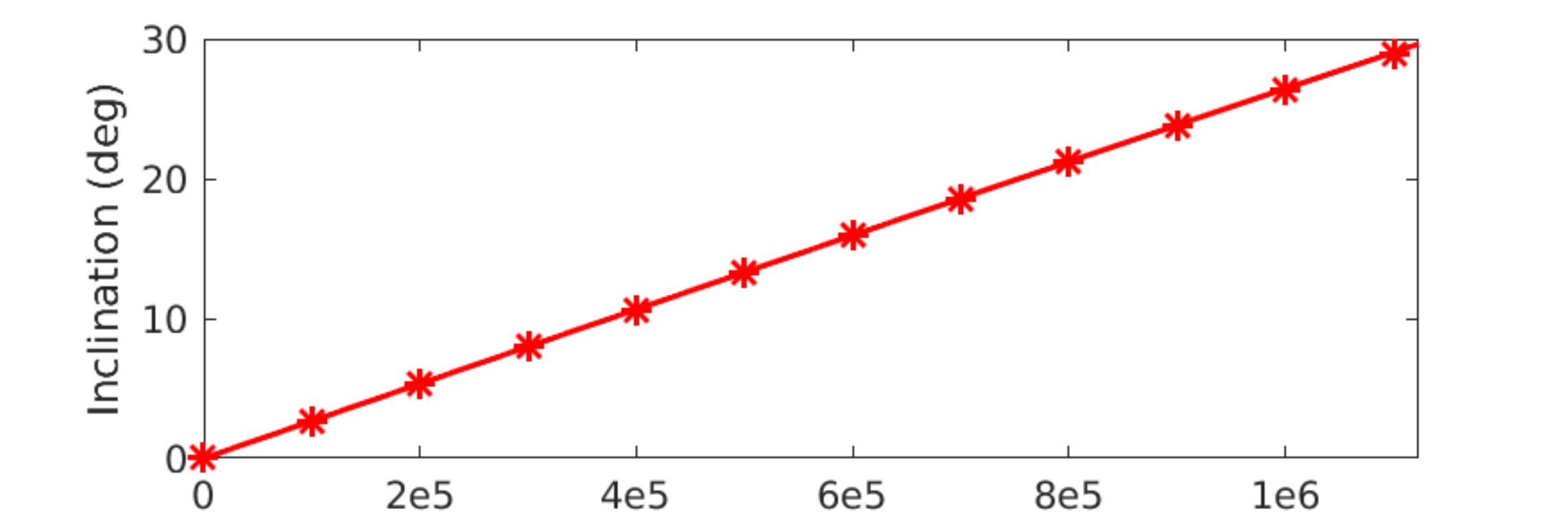}
	\includegraphics[width=\columnwidth]{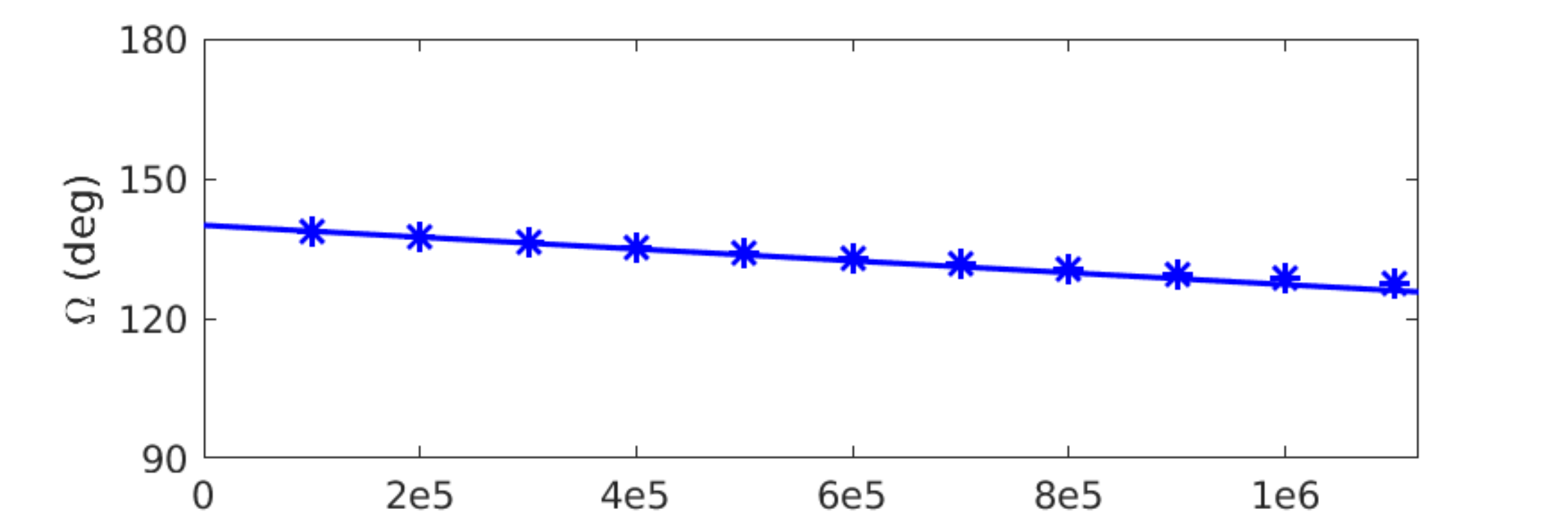}
	\includegraphics[width=\columnwidth]{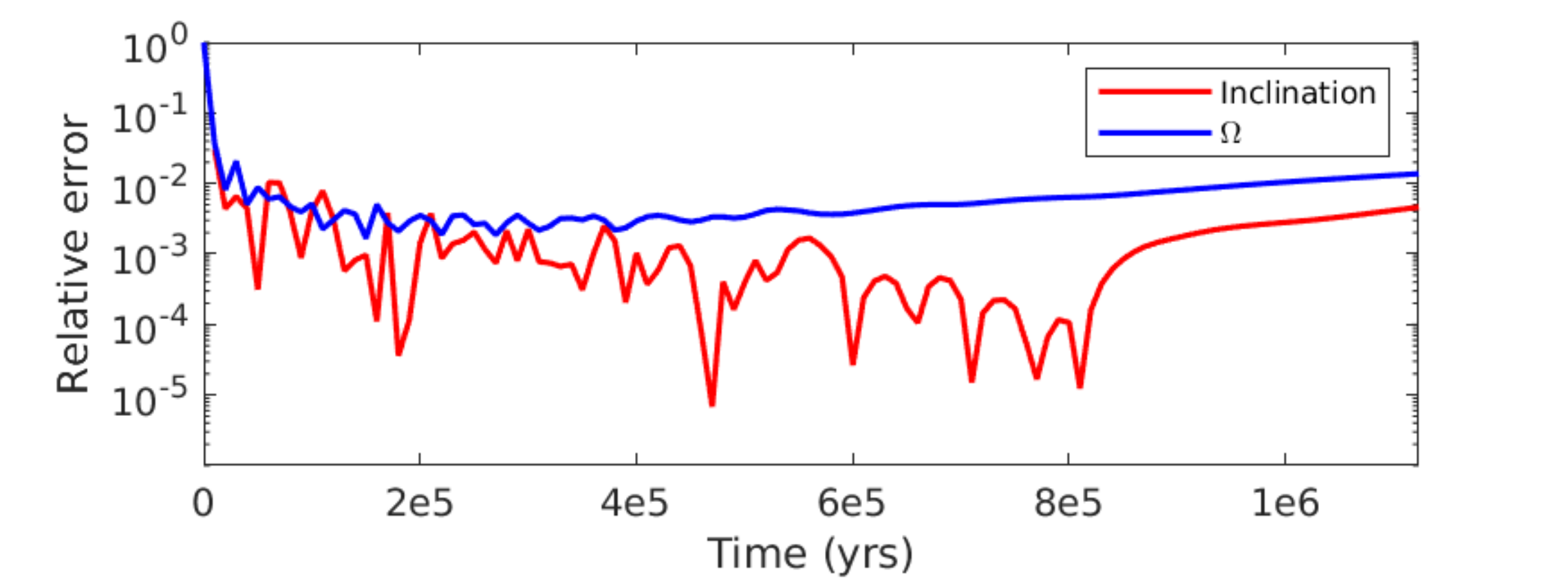}
	\caption{Comparison between the interpolated formulas~\eqref{i_d}-\eqref{Om_d} (solide line) and the numerical simulation of Hamiltonian~\eqref{Hamiltonian} (stars) for the evolution of a disk whose mass decreases exponentially and which is perturbed by a wide binary companion. The top panel shows the inclination of the disk while the middle panel shows the longitude of the ascending node. The relative errors between the two formulations are displayed in the bottom panel.}
	\label{validation_inter}
\end{figure}

To test the precision of our approximate formulas, we show a comparison of the evolution of the disk when using the Hamiltonian formulation \eqref{Hamiltonian} and our approximation~\eqref{i_d}-\eqref{Om_d}, in the case of an exponentially decreasing disk mass. The following system parameters are considered for the simulation: $m_0=m_B=1m_\odot$, $a_{\rm in}=2$ au, $a_{\rm out}=30$ au, $a_B=1000$ au, $i_B=60^\circ$, $\Omega_B=50^\circ$, an initial disk mass of $8m_{\rm Jup}$, a disk surface profile density proportional to $r^{-0.5}$, and $T_0=2.8\times 10^5$ yr. The evolution of the inclination and the longitude of the ascending node of the disk is displayed in Fig.~\ref{validation_inter} (top panels), with stars representing the formulation \eqref{Hamiltonian}, while solid lines show our approximation~\eqref{i_d}-\eqref{Om_d}. The two models are in good agreement, as shown by the relative error displayed in the bottom panel (relative error below $1\%$ during the whole disk lifetime).

To summarize, in this section we numerically found formulas for the evolution of the inclination (Eq.~\eqref{i_d}) and longitude of the ascending node (Eq.~\eqref{Om_d}) of a rigidly precessing disk perturbed by a distant binary companion (expressed with respect to the initial plane of the disk). 



\subsubsection{Analytical validation of the precession frequency}
\label{sec:analytic}

The frequency obtained in Eq. \eqref{freq_disk} can also be derived analytically, and thus provide a check of its validity. To do so, we assume that the disk can be represented by a continuum of concentric circular annuli. It is useful to introduce the complex inclination $W$, such as $W(r,t)=i(r,t) \exp{(\mathrm{i} \Omega)}$. The disk is warped if $i$ varies with radius, and twisted if $\Omega$ varies with radius. $W$ relates to the variables $p$ and $q$ of Eq.~\eqref{eq:pq} as 
$\text{Re}(W) = q$ and $\text{Im}(W) = p$. Similarly, we represent the binary star by its complex inclination $W_{\rm B}=i_{\rm B} \exp{(\mathrm{i} \Omega_{\rm B})}$. 
The disk extends from $a_{\rm in}$ to $a_{\rm out}$. We assume that the disk surface density goes as
\begin{equation}
\Sigma(r)=\Sigma_0\left(\frac{r}{a_{in}} \right)^{-p}
\end{equation}
and follows a Keplerian rotation around the central star of mass $m_0$, with a Keplerian frequency $n$. \\

A massive companion (in our case a binary star) exerts an external torque onto the disk. If we consider the disk to be a continuum of annuli with radius $r$, the secular interaction between each of these annuli and the binary star companion is equivalent to that described by the Laplace-Lagrange theory as seen in Section \ref{sec:approx}. The equation for an annulus of the disk is \citep{Lubow_2001}:
\begin{align}
\Sigma r^2 n \frac{\partial W}{ \partial t} &=\mathrm{i} G m_B \Sigma \mathcal{K}~ (W_{\rm B}-W).  \label{eq:Wdb}
\end{align} 
Here $\mathcal{K}$ is a symmetric function related to the well-known Laplace coefficients, defined by
\begin{equation}
\mathcal{K}(r,r')=\frac{r r'}{4 \pi}\int_0^{2\pi}\frac{\cos\varphi~\mathrm{d}\varphi}{\left(r^2+r'2-2rr'\cos\varphi \right)^{3/2}}.
\end{equation}
One can show that
\begin{equation}
\mathcal{K}(r,a_B)= \frac{1}{4a_B}\left(\frac{r}{a_B}\right)b_{3/2}^{(1)}.
\end{equation}
At first order in $r<a_B$, one has $b_{3/2}^{(1)}=3r/a_B$.

In order to get the precession of the whole disk, we multiply Eq.~\eqref{eq:Wdb} by $2\pi r$ and integrate over the radial extent of the disk. We assume that internal physical processes, such as pressure and self-gravity, maintain the disk rigid nodal precession, against the differential precession induced by the binary. In other words, $W$ does not depend on $r$. We have seen in the previous section how self-gravity can provide a mean to maintain such coherence \citep[see also][]{Batygin_2011}, and various authors \citep[see e.g.,][]{Larwood_1996,Zanazzi_2018a} have explored how pressure-driven bending waves can have the same effect. We also assume that the disk does not exert a back reaction onto the binary, so that $W_{\rm B}=0$ (since the reference plane here is the orbital plane of the binary). For $r<a_B$, we find that
\begin{equation}
\dot{W} = - \mathrm{i} \frac{3}{4}\lambda \left(\frac{a_{\rm out}}{a_B}\right)^3\frac{m_B}{m_0}n_{\rm out} W,
\end{equation} 
where $n_{\rm out}$ is the Keplerian frequency at the outer edge of the disk, and we have introduced
\begin{equation}
\lambda = \frac{\frac{5}{2}-p}{4-p}\frac{1-\eta^{4-p}}{1-\eta^{5/2-p}},
\end{equation}
with $\eta\equiv a_{\rm in}/a_{\rm out}$.
For clarity, let us define the nodal precession frequency $\omega_{\rm db}$ as
\begin{equation}
\label{eq:wdb*}
\omega_{\rm db} = \frac{3}{4}\lambda \frac{m_B}{m_0} \left(\frac{a_{\rm out}}{a_B}\right)^3  n_{\rm out}.
\end{equation}
This expression is the same as the one obtained by \cite{Zanazzi_2018b}. Then the first-order linear ODE rewrites
\begin{equation}
\label{eq:Wode}
\dot{W} = -\mathrm{i} \omega_{\rm db} W.
\end{equation}
If the inclination of the disk at $t=0$ is denoted by $i_0$, its longitude of the ascending node by $\Omega_0$, and therefore the initial complex inclination of the disk by $W_0=i_0\mathrm{e}^{\mathrm{i} \Omega_0}$, the solution of Eq.~(\ref{eq:Wode}) is
\begin{equation}
W(t) = i_0 \mathrm{e}^{\mathrm{i} (\Omega_0-\omega_{\rm db}t)}.
\label{id_anal}
\end{equation}
This implies that the inclination of the disk remains constant (since $
|W|=i_0$) and the longitude of the ascending node of the disk precesses as
\begin{equation}
\label{eq:omegat}
\Omega(t)=\Omega_0 - \omega_{\rm db}t.
\end{equation} 
For the parameters considered in the previous section, namely $m_0=1m_\odot$, $p=0.5$, $a_{in}=2$ au, and $a_{out}=30$ au, we find that $\omega_{\rm db}=2.7046$,
which we can compare to the value of the numerical coefficient of Eq.~\eqref{freq_disk}: $2 \times 2\pi \times 0.2145 = 2.695$. Note that the frequency has to be doubled since the sinus is in absolute value and we also multiply by a factor $2\pi$ to make the link between the ordinary frequency in Eq.~\eqref{freq_disk} with the angular frequency from Eq.~\eqref{eq:wdb*}. The agreement between the two values is very good. Note also that the influence of $a_{\rm in}$ is somewhat limited in Eq.~\eqref{eq:wdb*}, which explains that we did not notice a significant influence of $a_{\rm in}$ on the disk evolution in the numerical simulations of Section~\ref{sec:approx}.

\begin{table}
	\caption{Initial parameters for the simulations. The elements are expressed with respect to the initial plane of the disk.}
	\begin{tabular}{lrrr}
		\hline
		& Central star & Planet & Binary companion\\
		\hline
		mass  & $1$ $m_{\odot}$ & $U[1;5]$ $m_{\rm Jup}$ & $1$ $m_{\odot}$\\
		$a$ (au)  & & $20$ & $1000$\\
		$e$  & & $U[0.001;0.01]$ & $10^{-3}$, $0.1$, $0.3$, $0.5$\\ 
		$i$ ($^\circ$) & & $U[0.01;0.1]$ & $10^{-3}$, $10$, $20$, $30$, $40$,  \\
		& & & $50$, $60$, $70$ \\
		$\Omega$ ($^\circ$) & & $10^{-3}$& $10^{-3}$\\
		$\omega$ ($^\circ$) & & $U[0;360]$& $10^{-3}$\\
		$M$ ($^\circ$) & & $10^{-3}$ & $10^{-3}$\\
		\hline
	\end{tabular}
	
	\label{Body_param}
\end{table}

\section{Simulations}
\label{results}
In this section, we study the influence of the two additional effects described in the previous section, namely the disk gravitational potential (hereafter, GP) and the disk nodal precession caused by a binary companion (hereafter, NP), on the evolution of a migrating giant planet in a S-type binary system. This work pursues the study of \cite{Roisin_2020}, where such effects were not considered.

\begin{figure}
	\includegraphics[width=\columnwidth]{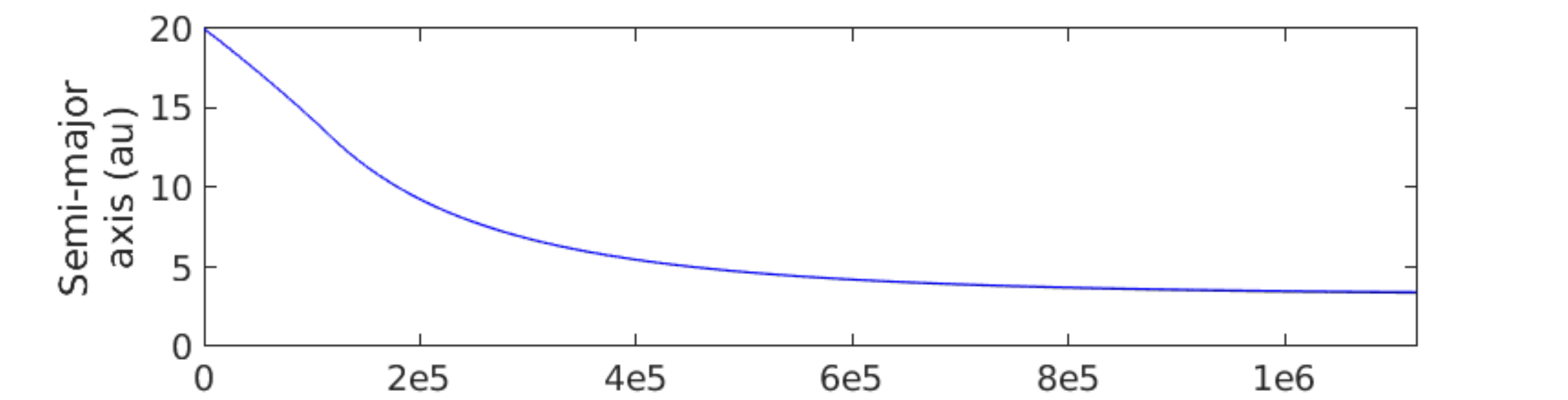}
	\includegraphics[width=\columnwidth]{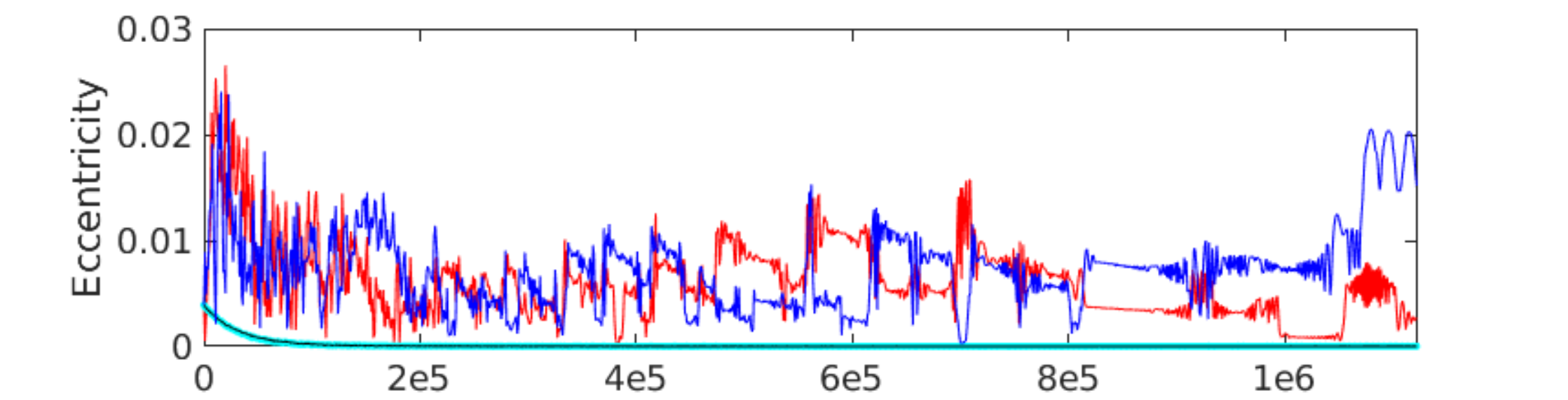}
	\includegraphics[width=\columnwidth]{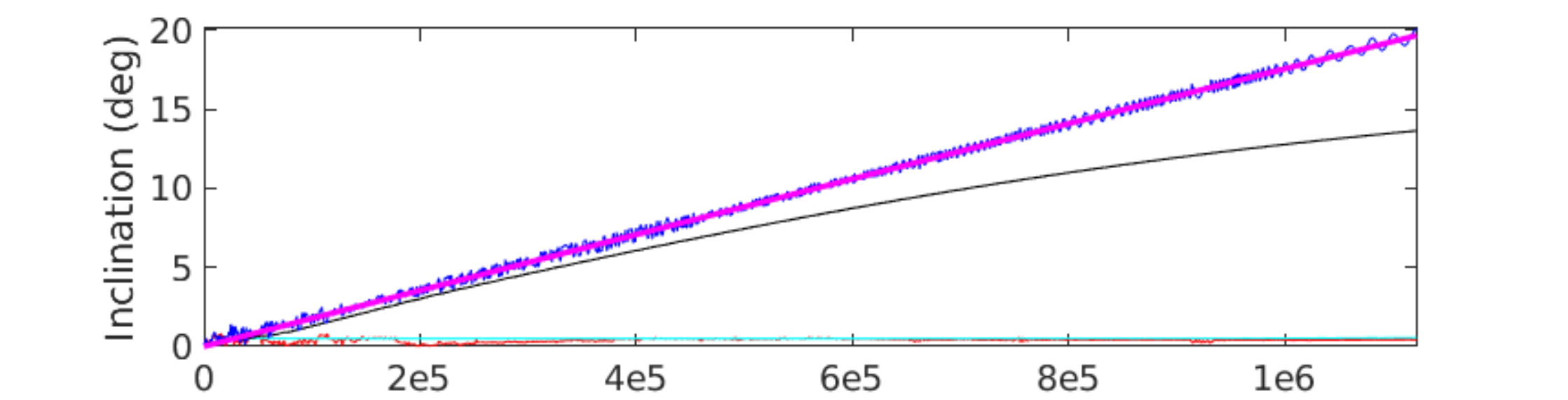}
	\includegraphics[width=\columnwidth]{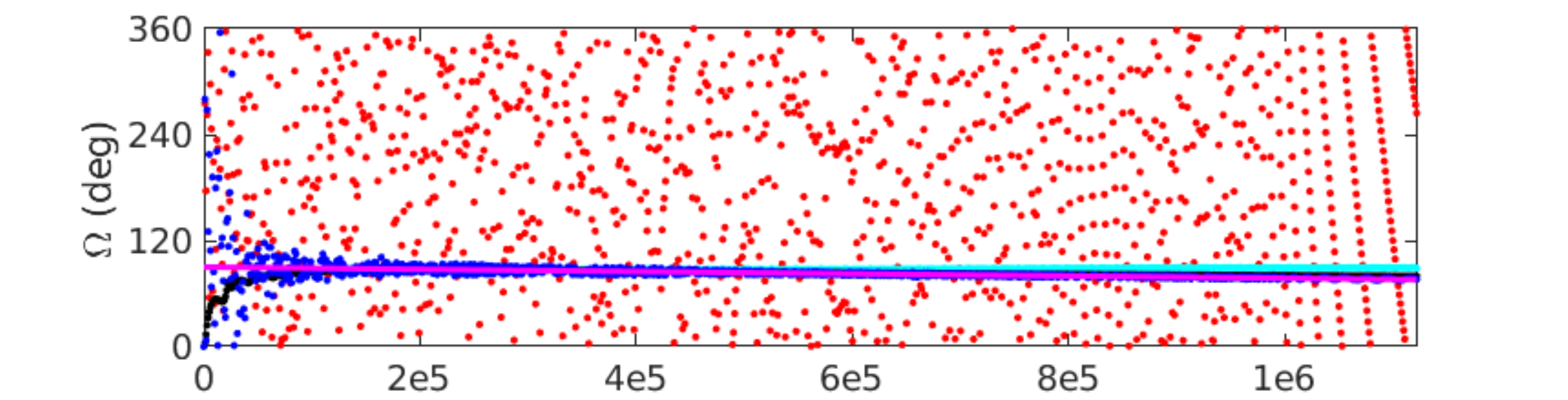}
	\includegraphics[width=\columnwidth]{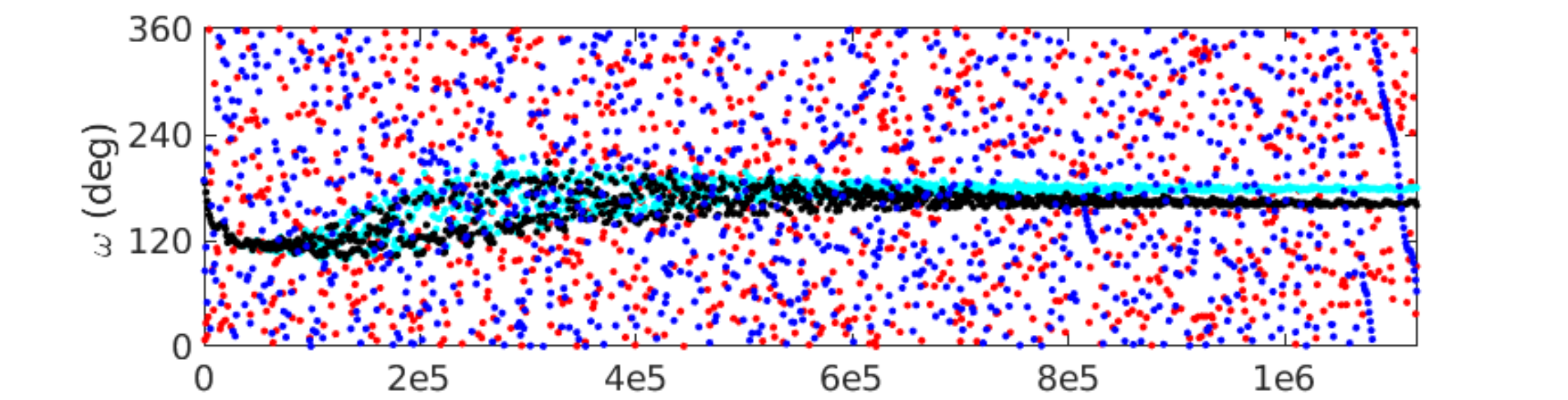}
	\includegraphics[width=\columnwidth]{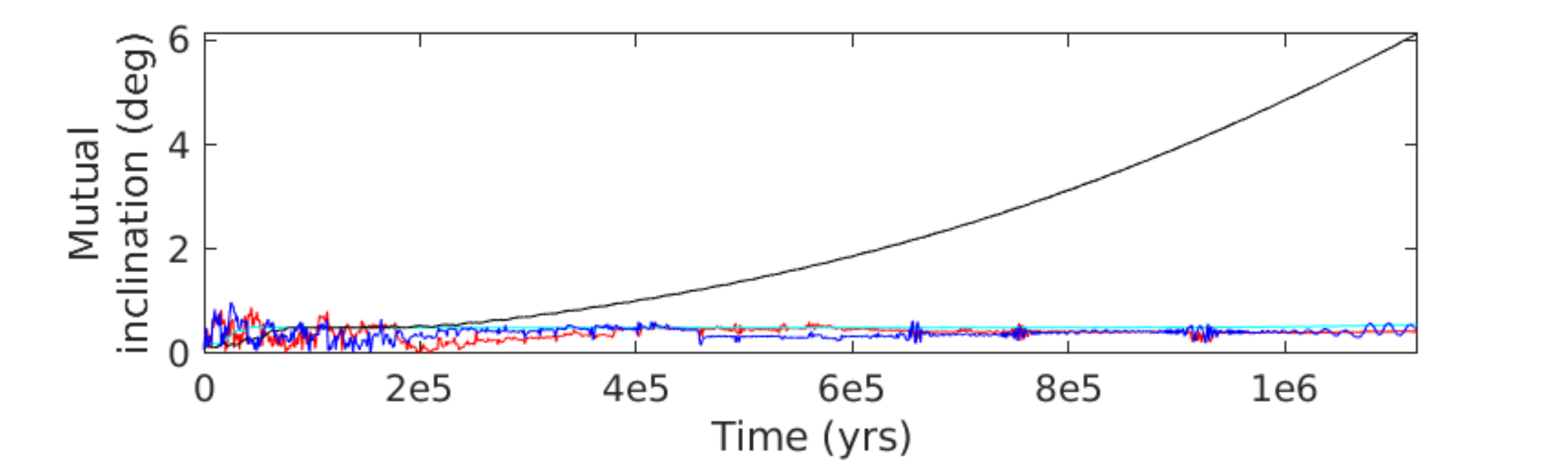}
	\caption{Typical evolution of a migrating giant planet in the presence of a distant binary companion when GP and NP effects are considered (blue curve). All elements are measured with respect to the initial disk plane. The cyan curves present the system evolving without GP/NP, the red curves with GP only, and the black lines with NP only. The magenta lines represent the disk evolution predicted through the formulas~\eqref{i_d}-\eqref{Om_d}. Initial parameters are $e_B=0.3$, $i_B=40^\circ$, $m_{pl}=1.53 m_{\rm Jup}$, and $\omega_{pl}=188^\circ$.}
	\label{fig:typical}
\end{figure}

\begin{figure}
	\includegraphics[width=\columnwidth]{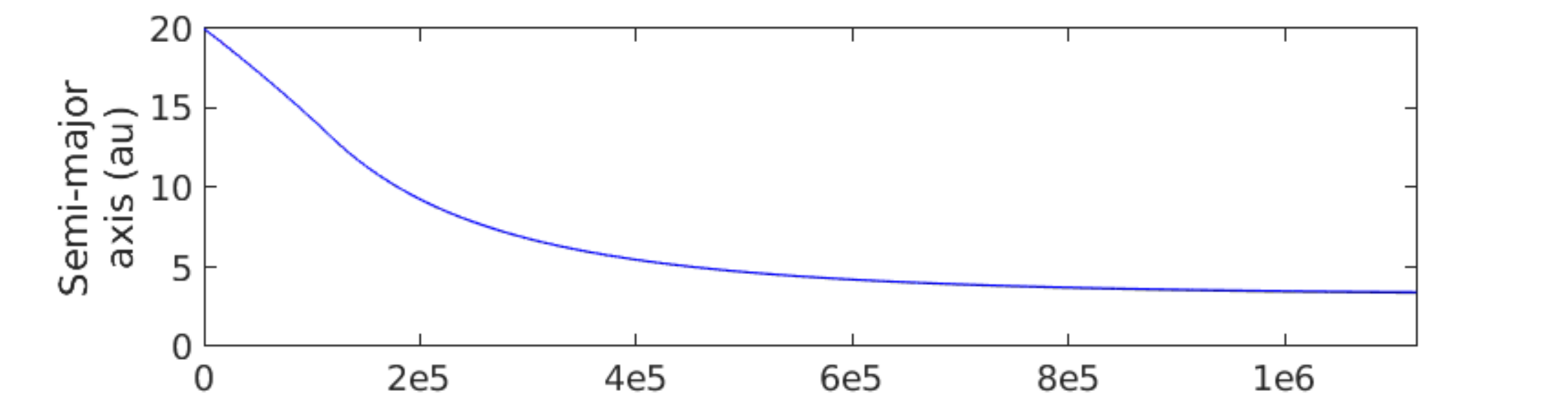}
	\includegraphics[width=\columnwidth]{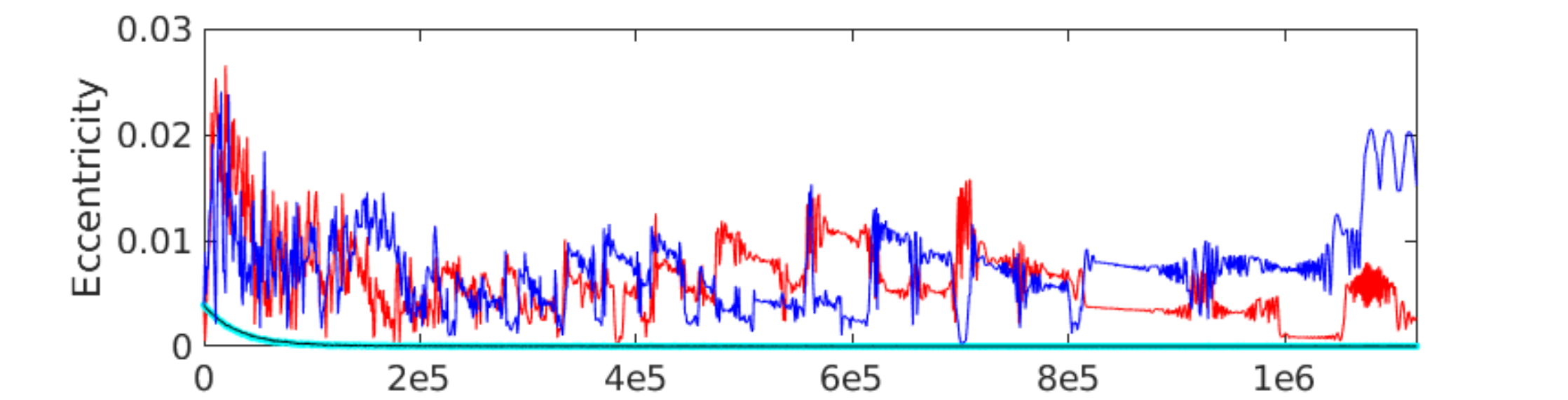}
	\includegraphics[width=\columnwidth]{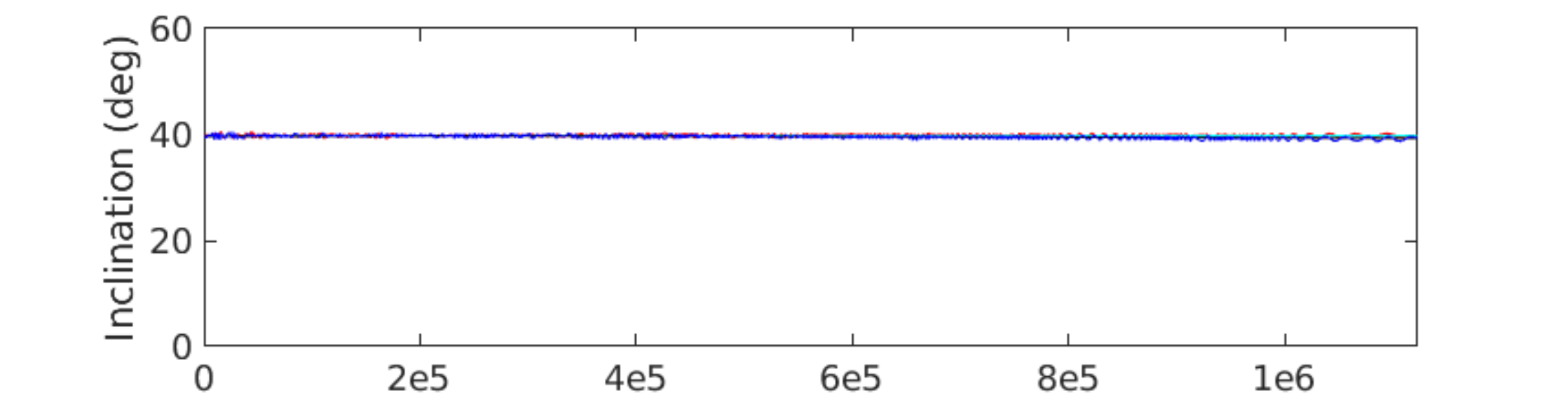}
	\includegraphics[width=\columnwidth]{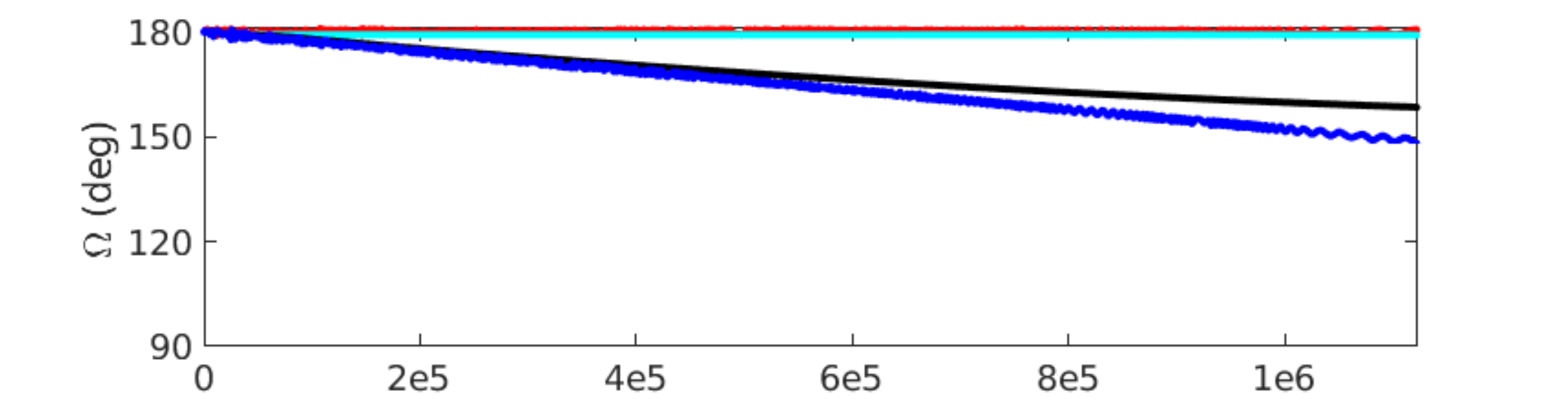}
	\includegraphics[width=\columnwidth]{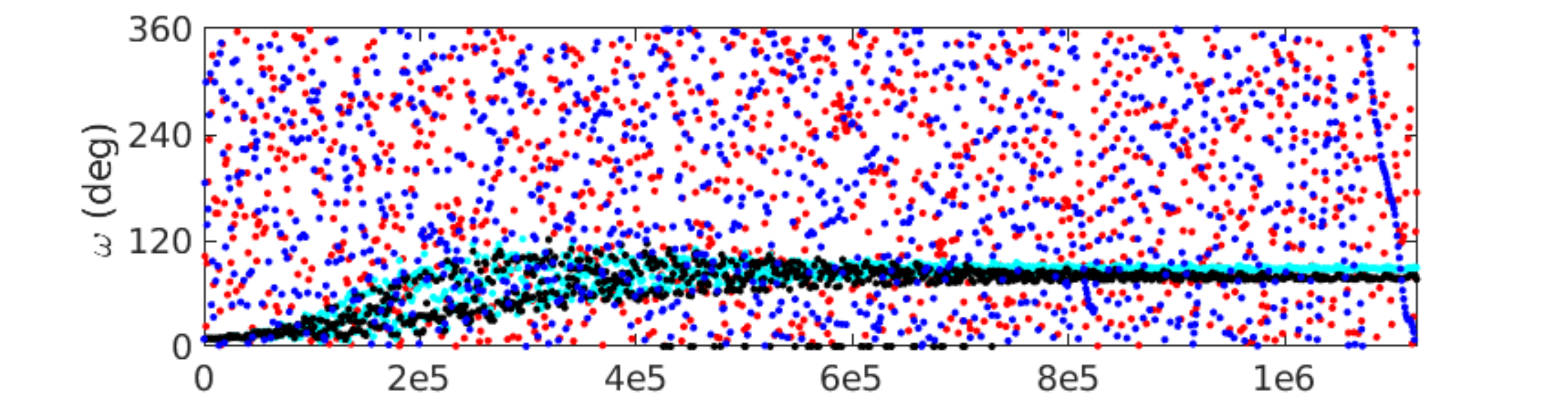}
	\includegraphics[width=\columnwidth]{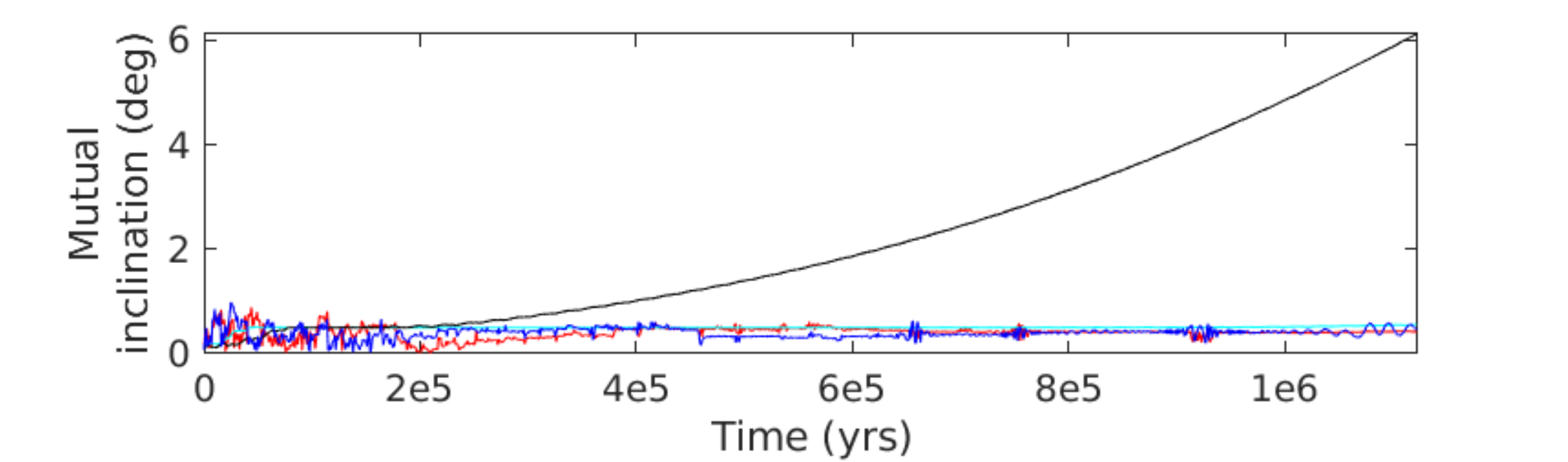}
	\caption{Same as Fig.~\ref{fig:typical}, but with respect to the invariant Laplace plane.}
	\label{fig:typicalLaplace}
\end{figure}

\subsection{Parameter set-up}
In this work, we carried out 3200 numerical simulations of the evolution of a single giant planet migrating in a S-type wide binary system. As in the previous section, the wide binary companion is located at a distance of $1000$ au. Regarding the disk parameters, as previously discussed in Section~\ref{sec:nbody}, we fix the disk inner edge to 0.05 au, the outer edge to 30 au, the surface profile density proportional to $r^{-0.5}$, and the disk dispersal time to $\sim~1$ Myr. Table~\ref{Body_param} summarizes the mass and orbital parameters considered for the two stars and the giant planet. For the binary companion, we used 8 different inclinations ranging from nearly $0^\circ$ to $70^\circ$, as well as 4 different eccentricities ($10^{-3}, 0.1, 0.3$, and $0.5$). For each combination of eccentricity and inclination values of the binary companion, we randomly drew 100 different initial conditions for the planet by considering uniform distribution for the planetary mass ($m_{pl}\sim U[1;5] m_{\rm Jup}$), eccentricity ($e_{pl}\sim U[0.001;0.01]$), inclination ($i_{pl}\sim U[0.01^\circ;0.1^\circ]$), and argument of the pericenter ($\omega_{pl}\sim U[0^\circ;360^\circ]$). 

The 3200 simulations were run with and without the GP and NP effects, for comparison.  The systems were followed for $5\times 10^8$~yr, with a time step of $5\times10^{-3}$~yr during the disk phase and $10^{-2}$~yr after the dissipation of the disk.

\subsection{Typical example}
We first present a typical example of the evolution of a migrating giant planet perturbated by a wide binary companion on an eccentric (0.3) and inclined ($40^\circ$) orbit. Fig.~\ref{fig:typical} displays the time evolution of the orbital elements of the planet during the disk phase for four different models: when the GP and NP are both included (blue curves), when the GP only is included (red curves), when the NP only is included (black curves), and finally when the GP and NP are not considered (cyan curves). 

While the planetary semi-major axis decreases due to the Type-II migration, the GP acting on the migrating planet tends to excite the planetary eccentricity and thus compete with the eccentricity damping associated with the Type-II migration. We can easily observe the gradual lowering of the damping effect associated to the decrease of the disk mass. Regarding the inclination of the planet, when both the GP and NP effects are present (blue curves), the planet closely follows the evolution of the disk, whose inclination and nodal precession computed from Eqs.~\eqref{i_d}-\eqref{Om_d} are indicated in Fig.~\ref{fig:typical} with magenta curves. The mutual inclination between the planet and the disk is also shown in the bottom panel and confirms that the planet stays in the disk during the whole disk phase when both effects are considered. Whith NP only included (black curve), the disk and the planet are gravitationally decoupled and precess at different rates in the binary plane.


In Fig.~\ref{fig:typicalLaplace}, the same system evolution is shown in the invariant Laplace plane. In this reference plane, the Lidov-Kozai resonance \citep{Lidov_1962, Kozai_1962} can be easily identified from the libration of the planetary argument of the pericenter. We observe that, when the GP is not considered, the argument of the pericenter rapidly librates around $90^\circ$, which indicates a capture of the planet in a Lidov-Kozai resonant state with the binary companion. On the contrary, the GP prevents the planet to be locked into the Lidov-Kozai resonance during the disk phase. This example clearly shows that GP is the dominant effect acting on the dynamics of a migrating planet when compared to the nodal precession, as previously envisioned in the discussion of \cite{Roisin_2020}. 


\begin{figure}
	\includegraphics[width=\columnwidth]{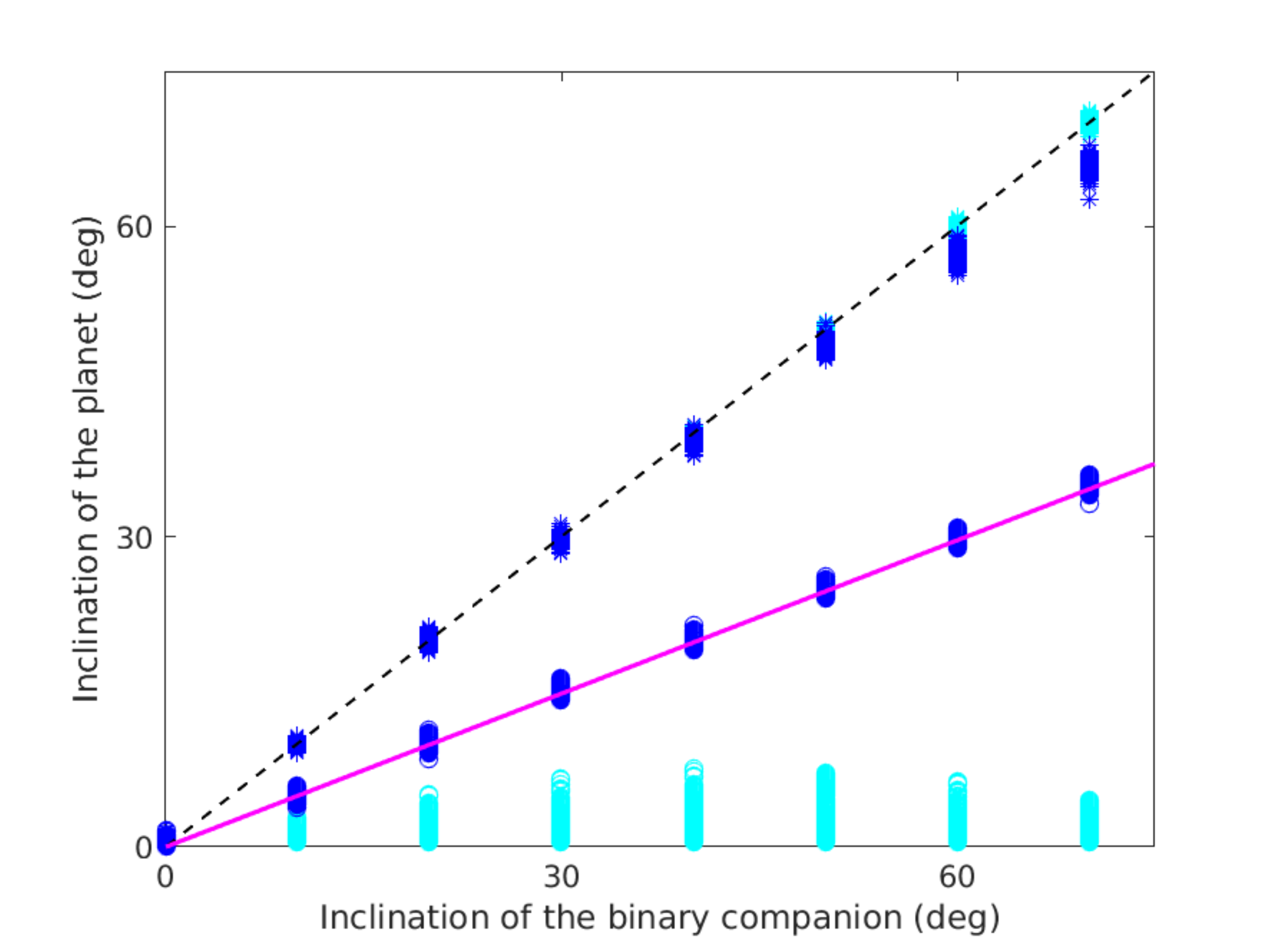}
	\caption{Planetary inclination versus inclination of the binary companion, at the dispersal of the disk. The blue (cyan) color refers to simulations with (without) GP/NP. The circle symbol refers to inclinations in the initial disk plane, and the star symbol in the invariant Laplace plane.}
	\label{i}
\end{figure}

\begin{table}
	\centering
	\caption{Percentages of the systems ending up in a Lidov-Kozai resonant state, for the different inclinations of the binary companion $i_B$.}
	\begin{tabular}{rrr}
		$i_B$ ($^\circ$)  & Without GP/NP (\%) & With GP/NP (\%)\\
		\hline
		$\le 30$ & 0 & 0 \\
		40 & 4.5 & 2.75 \\
		50 & 41.5 & 28.75\\
		60 & 22.25 & 37.5 \\
		70 & 17.5 & 30.25 \\
	\end{tabular}
	
	\label{LK_perc}
\end{table}

\begin{figure*}
	\includegraphics[width=\columnwidth]{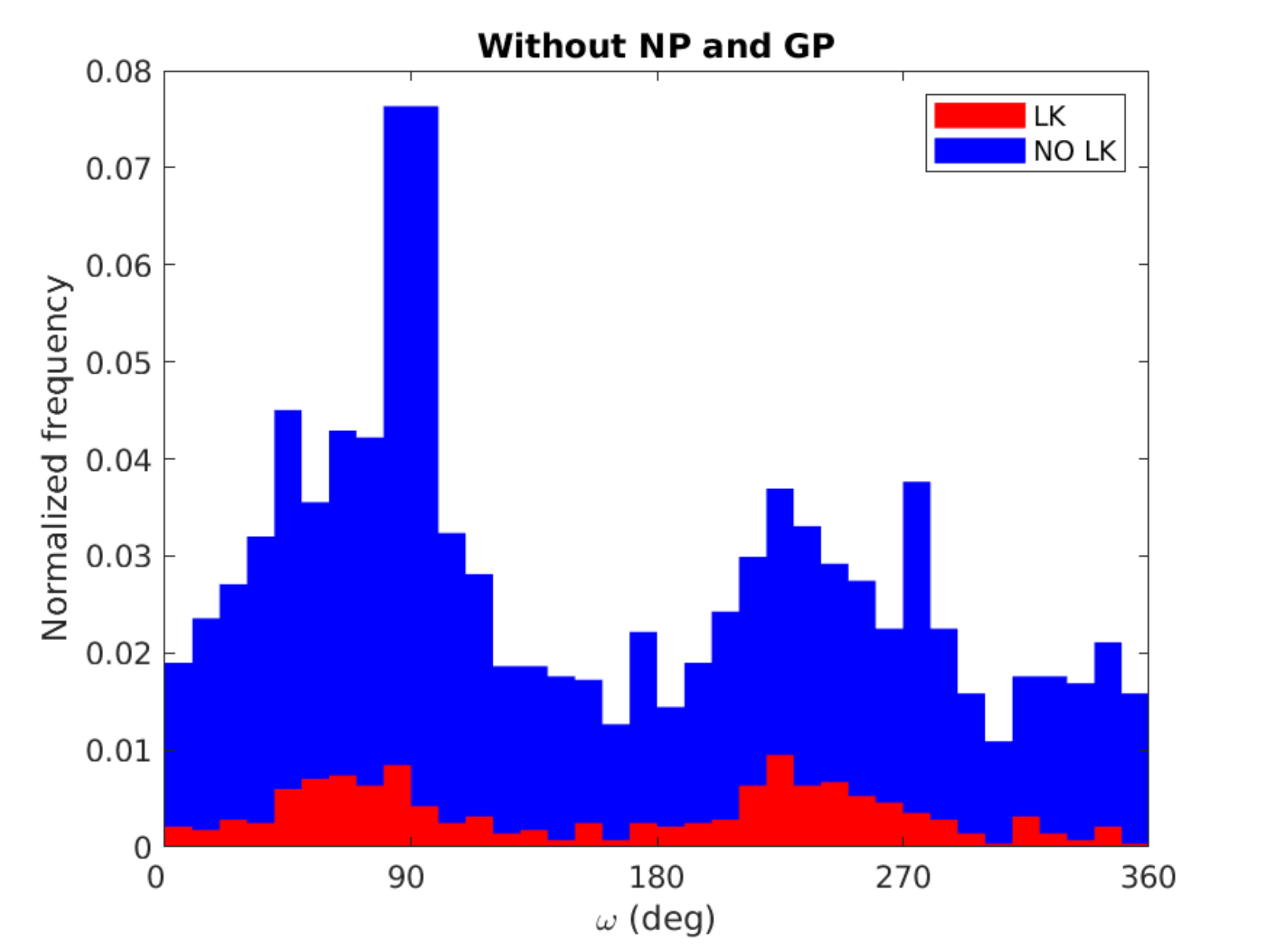}
	\includegraphics[width=\columnwidth]{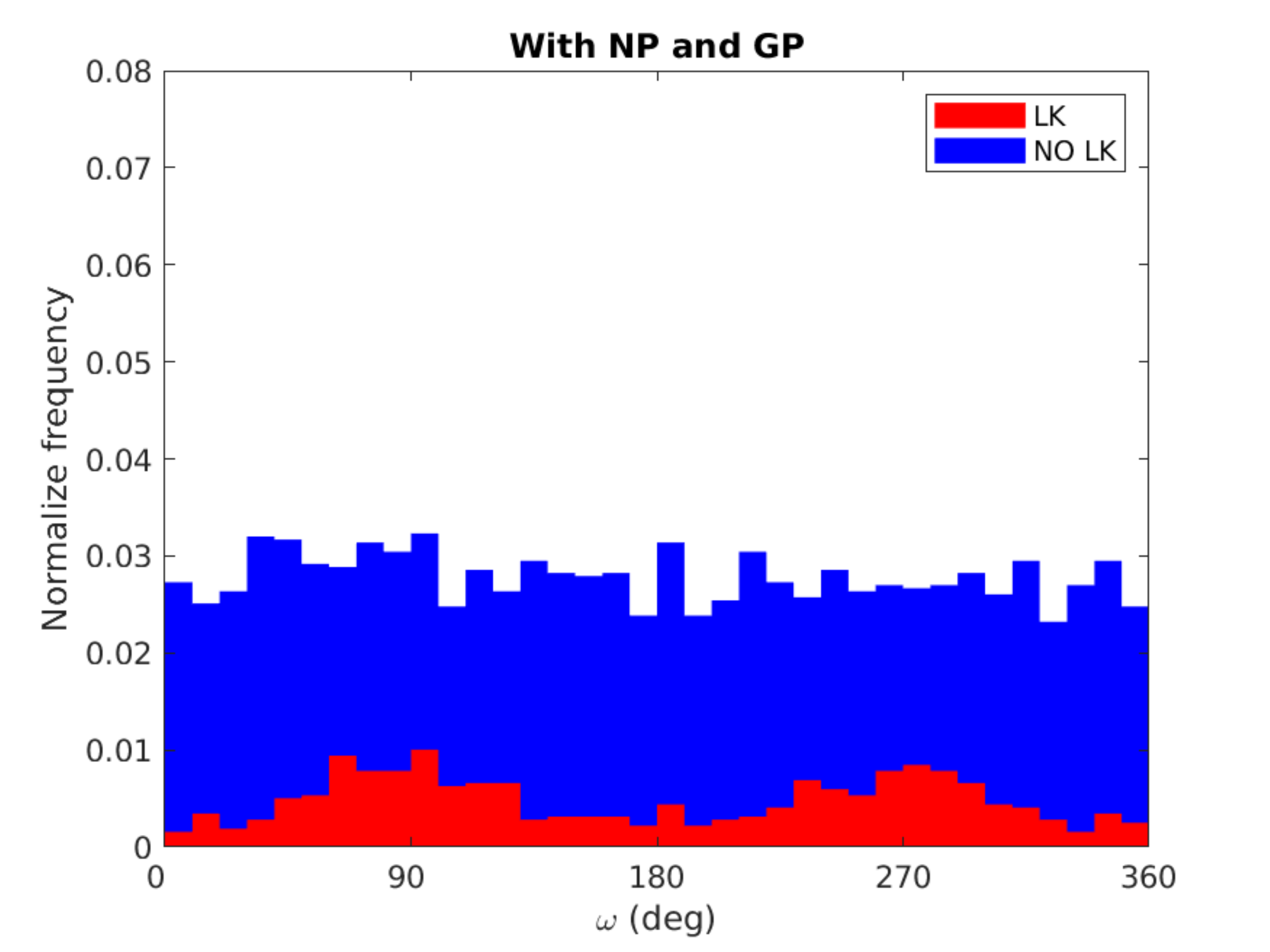}
	\caption{Normalized distribution of the pericenter argument of the planet (in the invariant Laplace plane reference frame) at the dispersal of the disk, for the simulations without GP/NP (left panel) and with GP/NP (right panel).}
	\label{hist_w_disk_abs}
\end{figure*}

\begin{figure*}
	\includegraphics[width=\columnwidth]{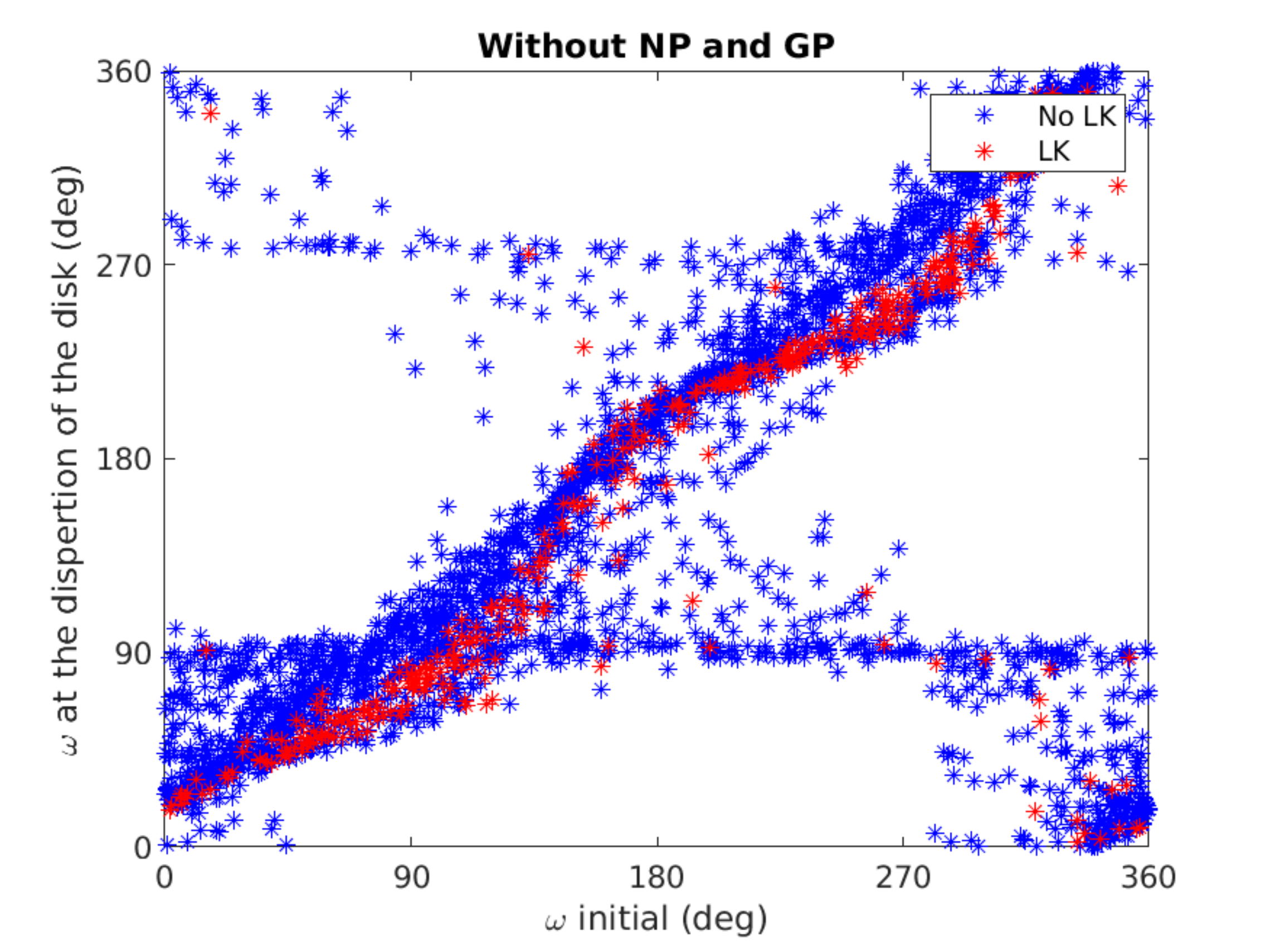}
	\includegraphics[width=\columnwidth]{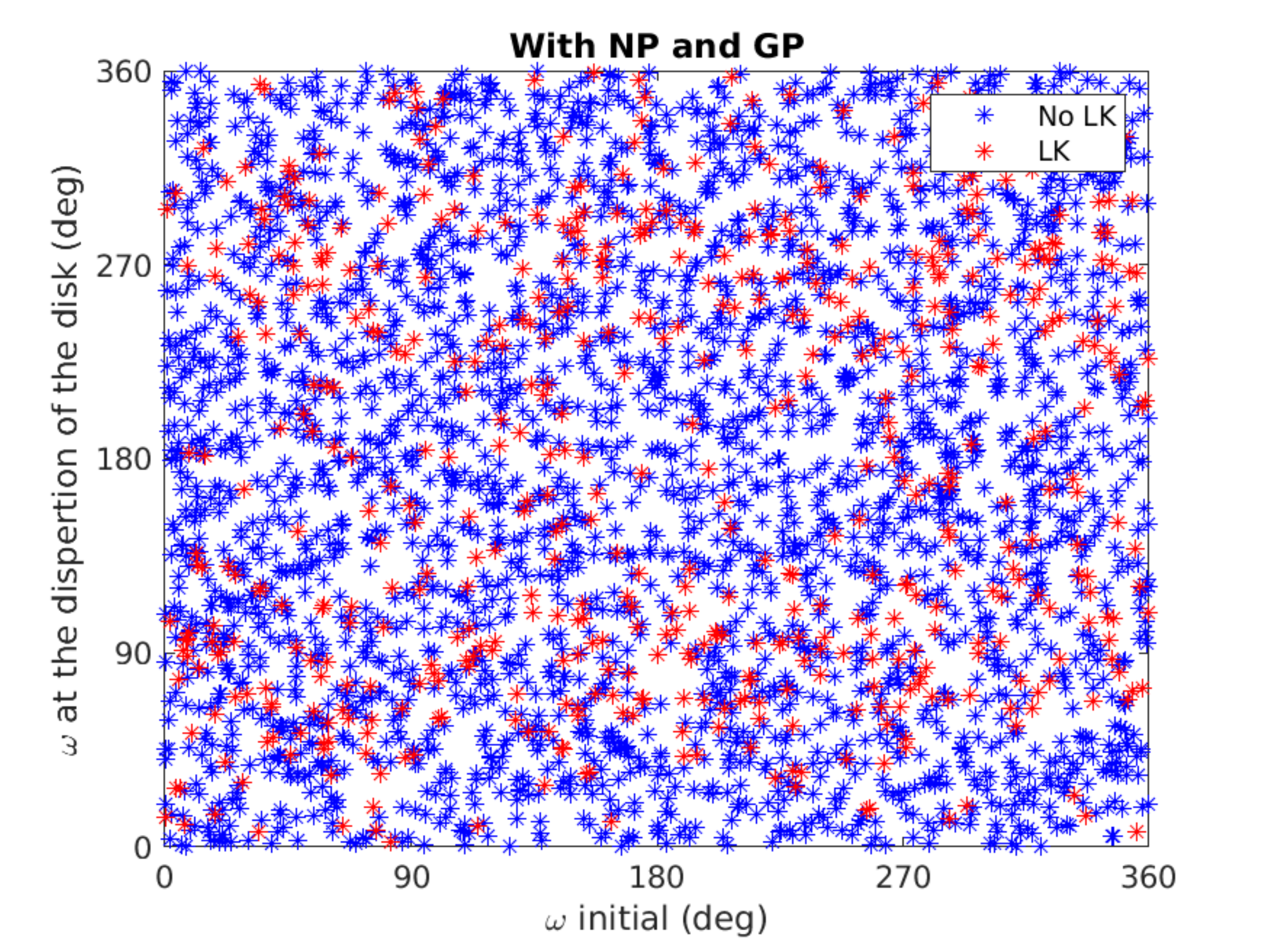}
	\caption{Pericenter argument of the planet at the dispersion of the disk versus initial pericenter argument (in the invariant Laplace plane reference frame), for the simulations without GP/NP (left panel) and with GP/NP (right panel).}
	\label{w_init_w_disk}
\end{figure*}

\begin{figure*}
	\includegraphics[width=\columnwidth]{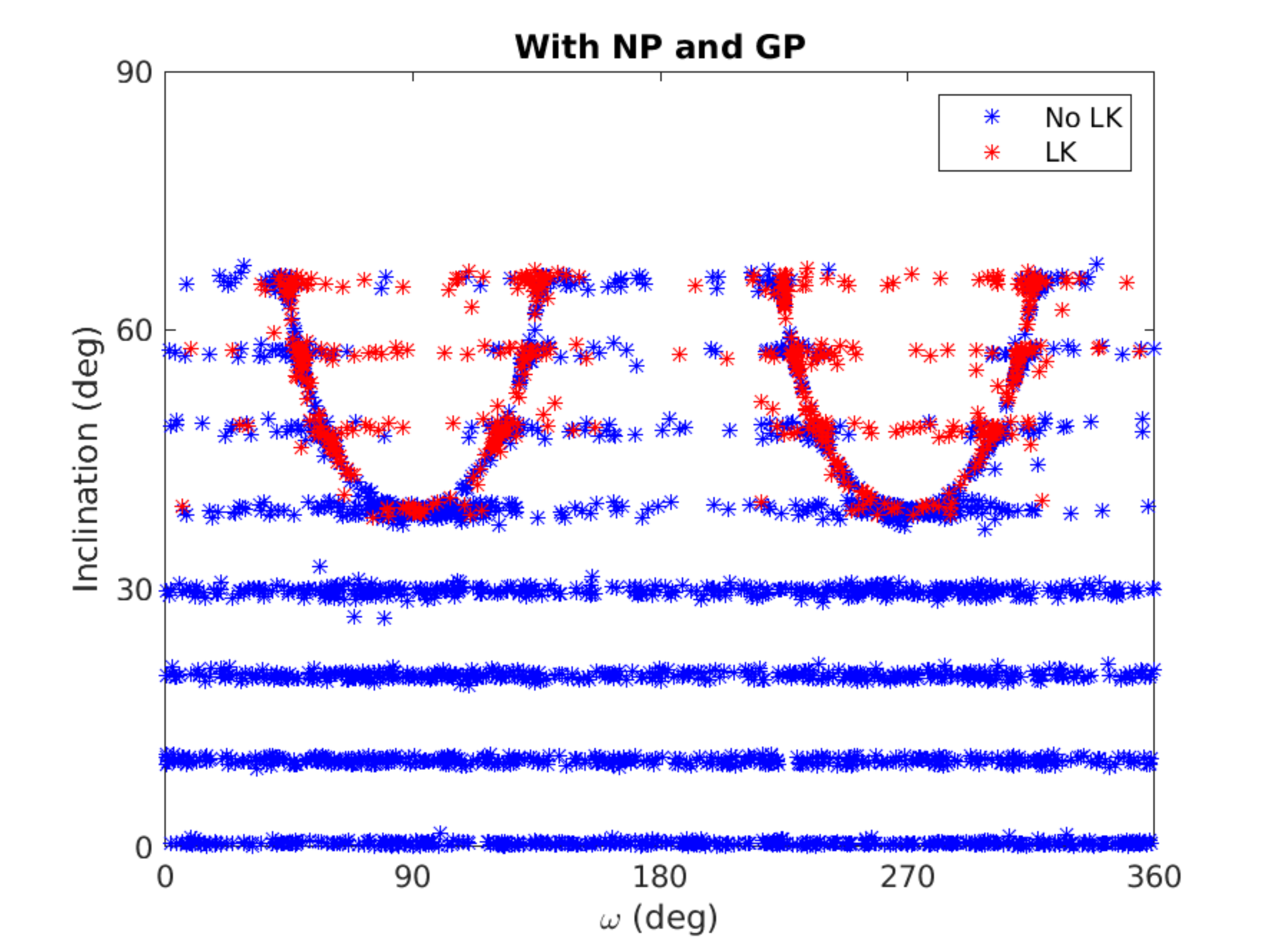}
	\caption{Inclination of the planet as a function of the argument of the pericenter (in the invariant Laplace plane reference frame), at the end of the simulation ($5\times 10^8$~yr).}
	\label{w_i}
\end{figure*}

\subsection{Results}

We now study the final parameters distribution of the 3200 simulations carried out in this work. In particular, to further study the GP and NP effects, we are interested in the planetary orbital values at the dispersal of the disk. In Fig.~\ref{i}, we report the inclination of the planet at the dissipation of the disk with respect to the initial inclination of the binary companion, with circle symbol indicating inclinations in the initial disk plane reference frame. As previously, the blue color refers to simulations with GP and NP, the cyan color to simulations without the two effects, and the magenta line to the disk inclination expected at the dissipation of the disk according to Eq.~\eqref{i_d}. We observe that almost all the planets follow this line when the GP and NP are included, which means that for all the initial parameters of the planet and the binary companion considered, the planet always stays in the disk during the disk phase, as also observed for the typical evolution in the previous section. Using star symbol, we report the inclinations with respect to the invariant Laplace plane. The black dashed line indicates when the planetary and binary inclinations in the Laplace plane are equal, highlighting the systems for which the mutual inclination stays almost constant during the disk phase, since most of the total angular momentum of the system is due to the binary companion. For simulations including GP and NP, we observe a small deviation from the black dashed line for highly inclined binary companions. This is due to the limitation of the Laplace-Lagrange theory used in Sect.~\ref{sec:approx} for high inclinations (see the discussion in the supplementary material of \cite{Batygin_2012} for more details).

A particular attention was also given to the Lidov-Kozai resonance. In \cite{Roisin_2020}, it was shown that the capture of the migrating planet in a Lidov-Kozai resonant state is far of being automatic when the binary companion is highly inclined. We first studied the percentage of systems ending up in a Lidov-Kozai resonance state (at the end of the simulation, i.e. $5 \times 10^8$ yr), for the different inclinations of the binary companion. The results are gathered in Table~\ref{LK_perc}. The higher the inclination of the binary companion, the higher the probability of capture in a Lidov-Kozai resonance state. Note that a planet is considered here as locked in a Lidov-Kozai resonance if its argument of the pericenter librates, in the Laplace reference frame, around $90^\circ$ or $270^\circ$ for at least three secular periods.

To further analyse the impact of the disk phase on the capture into a Lidov-Kozai resonance, we also focused on the values of the planet's argument of pericenter at the dispersal of the disk. In Fig.~\ref{hist_w_disk_abs}, we show the distribution of the pericenter argument of the planet (in the invariant Laplace plane) at the end of the disk phase, when considering the GP and NP effects (right panel) or not (left panel). The red color indicates that the planets are in a Lidov-Kozai resonant state at the end of the simulation. While accumulations of the pericenter argument around $90^\circ$ and $270^\circ$ during the disk phase are clearly visible without the GP and NP effects, these accumulations disappear when considering the two effects. As previously illustrated with the typical evolution shown in Fig.~\ref{fig:typical}, it is the GP effect of the disk acting on the planet that suppresses the Lidov-Kozai oscillations induced by the binary companion. The same result can be observed in Fig.~\ref{w_init_w_disk} where the pericenter argument of the planet at the dispersal of the disk is represented as a function of the initial pericenter argument value (in the invariant Laplace plane). On the left panel, when the GP and NP are not included in the simulations, we observe the accumulation to values close to $90^\circ$ and $270^\circ$, typical of the Lidov-Kozai resonance, during the disk phase. This particular shape completely disappears on the right panel where the GP and NP effects of the disk are present.

Once the disc has lost all of its mass, it becomes possible for the planet to fully undergo Lidov-Kozai oscillations driven by the binary, without those being quenched by the disk's GP. In Fig.~\ref{w_i}, we show the inclination of the planet with respect to the argument of its pericenter (in the invariant Laplace plane), at the end of the simulation (i.e., $5 \times 10^8$ yr). We see that although the GP and NP effects were considered during the disk phase, the arc-shape curves previously observed in \cite{Roisin_2020} and associated to the libration islands around the Lidov-Kozai equilibria are still present. As also shown in Table~\ref{LK_perc}, there are even slightly more planets evolving in the Lidov-Kozai resonance at the end of the simulations when the GP and NP effects are considered during the disk phase.

\section{Conclusions}
\label{conclusion}

In this work, we aimed to study the evolution of a planet migrating in the disk in the presence of a wide binary companion with separation of 1000 AU. We consider the self-gravity of the disk which efficiently maintains rigid precession throughout the disk perturbed by the wide binary companion. To do so, we derived new formulas of the evolution of a mass-decreasing disk under the influence of a wide binary companion in the initial disk plane reference frame. These formulas were obtained by splitting the disk in several massive rings adjacent to one another and evolving with the classical Laplace-Lagrange secular theory. We implemented this evolution in the $N$-body code presented in \cite{Roisin_2020} consisting in an adaptation of the well-known symplectic integrator SyMBA to wide binary systems and including the Type-II migration and eccentricity and inclination damping due to the disk. In this work, we also added the influence of the gravitational potential of the disk. Using this new code, we were able to accurately modelize the disc nodal precession caused by the binary companion, as well as take into account the gravitational force that the disk exerts on the planet. 

We observed that, while the nodal precession has a limited influence on 
the evolution of a giant planet migrating in a disk, the gravitational 
potential plays an important role in its evolution. In particular, 
alongside inclination damping forces, it acts to keep the planet in the 
disk and suppress the effect of the Lidov-Kozai resonance induced by the 
binary companion during the disk phase. Note that for closer-in binaries 
(100 au), \cite{Picogna_2015} found that the disc and the planet 
can be decoupled and large mutual inclinations between the two can be 
generated \citep[see also][]{Lubow_2016}. However in our case, the 
binary is further out and this effect was not observed in our simulations.
Regarding the establishment of the Lidov-Kozai resonance after the disk phase, one third of the planets with a highly inclined binary companion ended up locked in a Lidov-Kozai resonant state, that is roughly the same percentage as in \cite{Roisin_2020}, where the disk gravitational potential acting on the planet and the nodal precession induced by the binary companion on the disk were not considered. 

Finally, it is interesting to stress that, even for slightly inclined binary companions, a misalignment between the disk and the spin axis of the primary star is observed, due to the nodal precession induced by the binary companion (see Fig.~\ref{i_evolution_i_B}). \cite{Lai_2014} predicted that a primordial spin-orbit misalignment could be generated between a planet and its parent star in S-type binaries, although their model did not include a planet. Here we are able to confirm this prediction: since the gravitational potential and damping effect of the disk act to maintain the planet in the disk orbital plane during the disk-induced migration phase, a primordial spin-orbit misalignment could also be generated for circumprimary planets with an inclined binary companion.

\vspace{0.5cm}

\section*{Acknowledgements}
 The work of A. Roisin is supported by a F.R.S.-FNRS Research Fellowship and the work of J. Teyssandier is supported by a F.R.S.-FNRS Postdoctoral Research Fellowship. Computational resources have been provided by the Consortium des Equipements de Calcul Intensif, supported by the FNRS-FRFC, the Walloon Region, and the University of Namur (Conventions No. 2.5020.11, GEQ U.G006.15, 1610468 and RW/GEQ2016).

\section*{Data availability}
The data underlying this article will be shared on reasonable request to the corresponding author.

\bibliographystyle{mnras}
\bibliography{bibliography} 


\appendix

\section{Details about the change of coordinates between the orbital plane of the binary and the initial plane of the disk}
\label{Calculation_details}

In Section~\ref{sec:analytic}, we analytically determined the evolution of the disk's inclination and longitude of the ascending node with respect to the orbital plane of the binary. In this Appendix, we aim to explain the particularities highlighted in the formulas~\eqref{i_d} (i.e., the amplitude of the sinus function) and \eqref{Om_d} (i.e., the modulo $\pi$) of Section~\ref{sec:approx} regarding the evolution of the disk with respect to the initial disk plane reference frame.

The direction of the angular momentum of the disk can be expressed in the binary companion's orbital plane reference frame as \citep{Murray_1999}: 
\begin{equation}
	\begin{pmatrix}
		\sin i \sin \Omega(t) \\
		\sin i \cos \Omega(t) \\
		\cos i
	\end{pmatrix},
\end{equation}
with $i$ the inclination of the disk and $\Omega$ the longitude of the ascending node of the disk. Recall that the inclination is constant in this reference frame. Since our purpose is to express the angles with respect to the initial plane of the disk, we first apply a rotation around the $z$-axis with an angle equivalent to the initial longitude of the ascending node $\Omega_0$ and then a rotation around the $x$-axis with the inclination $i=i_0$. We find 
\begin{equation}
\begin{aligned}
	\begin{pmatrix}
	\sin i_d(t) \sin \Omega_d(t) \\
	-\sin i_d(t) \cos \Omega_d(t) \\
	\cos i_d(t)
	\end{pmatrix}
	&
	=
	\begin{pmatrix}
	1 & 0 & 0 \\
	0 & \cos i & -\sin i \\
	0 & \sin i & \cos i \\
	\end{pmatrix}
	\\
	&\hspace{-0.5cm}
	\begin{pmatrix}
	\cos \Omega_0 & \sin \Omega_0 & 0\\
	-\sin \Omega_0 & \cos \Omega_0 & 0 \\
	0 & 0 & 1
	\end{pmatrix}
	\begin{pmatrix}
	\sin i \sin \Omega(t) \\
	-\sin i \cos \Omega(t) \\
	\cos i
	\end{pmatrix},
\end{aligned}
\end{equation}
with $i_d(t)$ and $\Omega_d(t)$ the inclination and longitude of the ascending node of the disk, respectively, with respect to the initial plane of the disk. Using Eq.~\eqref{eq:omegat} and some trigonometrical identities, we obtain
\begin{align}
\begin{pmatrix}
 \sin i_d(t) \sin \Omega_d(t) \\
- \sin i_d(t) \cos \Omega_d(t) \\
 \cos i_d(t)
\end{pmatrix}
=
\begin{pmatrix}
\sin i \sin (\omega_{db} t) \\
- \cos i \sin i \left[1-\cos(\omega_{db} t) \right] \\
\frac{1}{2} \left[1+\cos(2i) + (1-\cos(2i))\cos(\omega_{db}t)\right]
\end{pmatrix}
\end{align}
The last component of the disk angular momentum vector ranges from $\cos(2i)$ to 1, which explains the amplitude for $i_d(t)$ of $2i_B$ found in Eq.~\eqref{i_d}. Considering that the inclination $i_d(t)$ varies between 0 and $\pi$ and since the expression of the second component of the vector is constant in sign, the variation of $\Omega_d(t)$ takes place on an interval of length $\pi$ instead of $2\pi$.


\bsp	
\label{lastpage}
\end{document}